\RequirePackage[2020-02-02]{latexrelease}
\documentclass[aps,pre,superscriptaddress,amsmath,amssymb,twocolumn]{revtex4}
\usepackage{float}
\usepackage[caption = false]{subfig}
\usepackage{graphicx}
\usepackage{amsmath}
\usepackage{hyperref}
\usepackage{color}
\usepackage{bigints}
\usepackage{mathrsfs}
\usepackage{algorithm}
\usepackage{algpseudocode}
\usepackage{multirow}
\usepackage{tabularx}
\usepackage[normalem]{ulem}
\usepackage{wrapfig}
\definecolor{darkblue}{RGB}{8,81,156}
\hypersetup{colorlinks,breaklinks,
	linkcolor=darkblue,urlcolor=darkblue,
	anchorcolor=darkblue,citecolor=darkblue}

\date{\today}

\allowdisplaybreaks

\definecolor{dark-purple}{RGB}{118,42,131}
\definecolor{dark-green}{RGB}{27,120,55}
\definecolor{light-purple}{RGB}{231,212,232}
\definecolor{LIGHT-PURPLE}{RGB}{194,165,207}
\definecolor{light-green}{RGB}{168,216,183}
\definecolor{gray}{RGB}{186,186,186}
\definecolor{super-dark-green}{RGB}{0,69,41}
\definecolor{super-dark-purple}{RGB}{63,0,125}
\definecolor{super-dark-blue}{RGB}{8,48,107}
\definecolor{super-dark-red}{RGB}{165,0,38}

\definecolor{super-dark-purple}{RGB}{64,0,75}
\definecolor{super-dark-green}{RGB}{0,68,27}

\usepackage{tabularx,booktabs} 

\usepackage{array}
\newcolumntype{L}[1]{>{\raggedright\let\newline\\\arraybackslash\hspace{0pt}}p{#1}}
\newcolumntype{C}[1]{>{\centering\let\newline\\\arraybackslash\hspace{0pt}}m{#1}}
\newcolumntype{R}[1]{>{\raggedleft\let\newline\\\arraybackslash\hspace{0pt}}m{#1}}

\setcitestyle{super}

\usepackage{natbib}
\usepackage{setspace}
\usepackage{esint}
\usepackage{amssymb} 

\newcommand{\R}{\mathbb{R}}

\usepackage[normalem]{ulem}

\definecolor{dark-purple}{RGB}{118,42,131}
\definecolor{dark-green}{RGB}{27,120,55}
\definecolor{light-purple}{RGB}{231,212,232}
\definecolor{LIGHT-PURPLE}{RGB}{194,165,207}
\definecolor{light-green}{RGB}{168,216,183}
\definecolor{gray}{RGB}{186,186,186}
\definecolor{super-dark-green}{RGB}{0,69,41}
\definecolor{super-dark-purple}{RGB}{63,0,125}
\definecolor{super-dark-blue}{RGB}{8,48,107}
\definecolor{super-dark-red}{RGB}{124,14,41}

\usepackage{versions}

\begin{document}

\allowdisplaybreaks


\title{Robust estimation of position-dependent anisotropic diffusivity tensors from stochastic trajectories}


\author{Tiago S. Domingues}
\affiliation{Department of Chemical and Environmental Engineering, Yale University, New Haven, CT 06520}
\author{Ronald Coifman}
\affiliation{Department of Mathematics, Yale University, New Haven, CT 06520}
\author{Amir Haji-Akbari}
\email[Corresponding Author:]{amir.hajiakbaribalou@yale.edu}
\affiliation{Department of Chemical and Environmental Engineering, Yale University, New Haven, CT 06520}


\date{\today}

\begin{abstract}
\noindent
\textbf{Abstract:}
Materials under confinement can possess properties that deviate considerably from their bulk counterparts. Indeed, confinement makes all physical properties position-dependent and possibly anisotropic, and characterizing such spatial variations and directionality has been an intense area of focus in experimental and computational studies of confined matter. While this task is fairly straightforward for simple mechanical observables, it is far more daunting for transport properties such as diffusivity that can only be estimated from autocorrelations of mechanical observables. For instance, there are well established methods for estimating diffusivity from experimentally observed or computationally generated trajectories in bulk systems. No rigorous generalizations of such methods, however, exist for confined systems. In this work, we present two filtered covariance estimators for computing anisotropic and position-dependent diffusivity tensors, and validate them by applying them to stochastic trajectories generated according to known diffusivity profiles. These estimators can accurately capture spatial variations that span over several orders of magnitude and that assume different functional forms. Our kernel-based approach is also very robust to implementation details such as the localization function and time discretization and performs significantly better than estimators that are solely based on local covariance. Moreover, the kernel function does not have to be localized and can instead belong to a dictionary of orthogonal functions. Therefore, the proposed estimator can be readily used to obtain functional estimates of diffusivity rather than a tabulated collection of pointwise estimates. Nonetheless, the susceptibility of the proposed estimators to time discretization is higher at the immediate vicinity of hard boundaries. We demonstrate this heightened susceptibility to be common among all covariance-based estimators. 
\end{abstract}


\maketitle

\section{Introduction}

\noindent
Confinement is known to alter the physicochemical properties of matter,\cite{AlcoutlabiJPhysCondMat2005} as confined materials can exhibit properties that deviate considerably from their bulk counterparts with identical compositions and under the same thermodynamic conditions.\cite{LevingerScience2002} Examples include changes in thermodynamic properties such as melting \cite{ChiashiACSNano2019} and boiling points,\cite{ChabanACSNano2012} and phase diagrams\cite{BinderSoftMtter2008, ChenPRL2016}, glass transition temperatures,\cite{ForrestPRL1996, SwallenScience2007} phase transition kinetics and mechanism,\cite{DurantGeophysResLett2005, ShawJPCB2005, AltabetPNAS2017,  HussainJACS2021} and mechanical\cite{KearnsAdvMater2010, GaoACSNano2020}, dielectric,\cite{FumagalliScience2018, OlivieriJPhysChemLett2021} and transport\cite{GuptaJChemPhys1997, HuPRL1991, PresslyMacromolecules2018, ZhangMacromolecules2018} properties. While such deviations can sometimes arise due to quantum effects,\cite{ParkApplPhysLett2001, HassanabadiNanoscale2020} they are usually a consequence of the broken translational symmetry of the underlying system due to confinement.\cite{HajiAkbariJChemPhys2014} Such symmetry breaking makes all thermodynamic, structural and transport properties position-dependent, with the latter also becoming anisotropic,~i.e.,~direction-dependent.\cite{SwanJChemPhys2011} For instance,  diffusivity, which is isotropic in most materials, becomes both anisotropic and position-dependent under confinement. Confinement is a ubiquitous means of tuning materials properties in systems as diverse as biological cells\cite{PinotCurrBiol2009, CarnesNatChemBiol2010} and colonies\cite{YouSciAdv2021} to semiconductors,\cite{HimmelbergerAdvFuncMater2013} solar cells\cite{ChenJACS2020} mesophases,\cite{FengNatMater2019} and heterogeneous catalysts.\cite{MitschkeChem2020} Understanding the interplay between  symmetry breaking and the subsequent spatial variations, and materials properties and function is not only fascinating from a fundamental perspective, but is also key for designing better materials for a wide variety of applications. As such, characterizing such spatial variations has been an intense area of focus in both experiments and simulations.

Since their inception,\cite{AlderJChemPhys1957, AlderJChemPhys1959} molecular simulations have been extensively utilized for studying confined states of matter, and for quantifying spatial variations in materials properties.\cite{AbrahamJChemPhys1978, SubramanianMolPhys1979, ToxvaerdJChemPhys1977, MagdaJChemPhys1985, HajiAkbariJCP2015} In the case of static properties,~i.e.,~mechanical observables that can be unambiguously estimated for a single configuration, this is a fairly trivial task and can be readily achieved via spatial binning. When it comes to  transport properties such as diffusivity, viscosity and heat and electric conductivity, however, such a task is more daunting, as these quantities can only be computed from autocorrelations of certain microscopic quantities.\cite{SearlesJChemPhys2000} In bulk systems, rigorous relationships exist between the autocorrelations of mechanical observables and the transport coefficient of interest (even if it is anisotropic) as long the corresponding linear constitutive relationships are applicable. While it is possible to construct localized forms of such autocorrelations, no rigorous relationship has been established between them and the local transport coefficient of interest. Consequently, previous efforts to characterize spatial variations in transport properties have often involved using heuristic approaches based on the assumption that such bulk relationships are valid at a local level.

A notable example is diffusivity. Assuming the validity of Fick's law, $\rho(\textbf{r},t|\textbf{r}_0,0)$, the conditional probability describing the spatiotemporal evolution of a typical point particle in the overdamped regime,
satisfies the \emph{Smoluchowski equation},\cite{DurangPhysRevE2013, VolpeRepProgPhys2016, BoJStatMechTheoryE2019}
\begin{equation}\label{Smoluch}
\frac{\partial\rho}{\partial t}=\nabla\cdot\Big[\textbf{D}(\textbf{r})\cdot\left[\nabla\rho+\beta\rho\nabla U(\textbf{r})\right]\Big]
\end{equation}
where  $U(\textbf{r})$ is an external (free) energy potential, and $\beta=1/k_BT$. Here, $\textbf{D}(\textbf{r})$ is a second-rank tensor that is a function of $\textbf{r}$ and  incorporates the anisotropy of diffusion. In a mean-field sense, the mobility statistics of any single point particle within most particulate systems is expected to follow Eq.~(\ref{Smoluch}) over reasonably long times. 

In systems with translational symmetry, both $U(\textbf{r})$ and $\textbf{D}(\textbf{r})$ are constant, and Eq.~(\ref{Smoluch}) simplifies into the Fokker-Planck equation,\cite{FokkerAnnPhys1914, PlanckSitzberPreussAkad1917} which has a closed-form analytical solution that constitutes the basis of the two classes of approaches used for estimating $\textbf{D}$ from single-particle trajectories.  The first approach is based on the Green-Kubo framework\cite{GreenJChemPhys1954, KuboJPhysSOcJpn1957} and relates diffusivity to the integral of the velocity autocorrelation function,\cite{LonguetHigginsJChemPhys1956} while the second approach-- also known as the Einstein relationship--\cite{EinsteinAnnPhys1905} is based on the Helfand framework\cite{HelfandPhysRev1960} and estimates $\textbf{D}$ from the asymptotic slope of the mean squared displacement (MSD) at $t\rightarrow\infty$. Under confinement, however, Eq.~(\ref{Smoluch}) no longer has a known closed-form analytical solution. Consequently, several computational efforts to estimate $\textbf{D}(\textbf{r})$ have employed \emph{ad hoc} estimators that are based on assuming local validity of the Einstein relationship and constructing local MSD functions for different regions of the confined system.\cite{MarrinkJPhysChem1994, TeboulJPhysCondMat2002, LanconPhysicaA2002, LiuJPhysChemB2004, DesaiJChemPhys2005, ShiJChemPhys2011} Apart from the fact that the validity of this core assumption is contested particularly in system where diffusivity undergoes large changes over short length scales, there is no intuitive way of defining such localized MSD functions in open systems.\cite{HajiAkbariJChemPhys2014}

There are, however, more sophisticated means of estimating position-dependent diffusivities. One such class constitutes Markov state models in which the spatial domain is partitioned into bins and a discretized version of the propagator of the  Smoluchowski time evolution restricted to each bin is computed. The local diffusivity can then be estimated as an expectation using the discretized propagator in the matrix form\cite{SicardJChemTheoryComput2021} or approximated from the entries of the transition matrix.\cite{HummerNewJPhys2005} This approach has, for instance, been utilized for estimating {\color{black}position-dependent} diffusivities in several systems such as {\color{black}hard spheres within a slit pore},\cite{MittalPhysRevLett2006}{\color{black} and small molecules within lipid bilayers.\cite{ComerJCTC2014, GhyselsJCTC2017, DeVosJCTC2018} Similar approaches have been used for estimating diffusivity profiles within a collective variable space.\cite{LjubeticJCP2014}}
While being more rigorous than the commonly used \emph{ad hoc} approaches, such methods have their own challenges, including their reliance on  optimal partitioning of the simulation box and proper selection of a propagation timescale. {\color{black}Another factor that limits their applicability to situations in which the diffusivity undergoes considerable changes throughout the system is the need to use a single propagation timescale.
}

It must be noted that the question of estimating position-dependent anisotropic diffusivity from particle trajectories belongs to an important class of problems in applied mathematics aimed at estimating the unknown parameters of a differential equation from its observed solution. While mostly overlooked in the molecular simulations community, such methods have widespread applications in areas as diverse as statistics, economics, finance and high-energy particle physics. One such class of methods are based on the observation that Eq.~\eqref{Smoluch} is a forward Kolmogorov equation, \cite{SchussTheoryStochastic2010} and, as such, its solution is equivalent to the probability distribution of particles evolving according to an associated stochastic differential equation (SDE). For instance, for isotropic diffusion, it is possible to use short-time asymptotic analysis to derive a likelihood function for short trajectories and to develop a Bayesian framework for estimating  diffusivity.\cite{noauthor_bayesian_nodate} Using such methods, however, require employing specialized algorithms that are fine tuned to the particular geometry of the corresponding system. Applying them to the question of estimating anisotropic diffusivity is therefore not trivial.

An important class of SDE-based methods are kernel-based estimators, which date back to the seminal works of Dacunha-Castelle and Florens-Zmirou,\cite{DacunhaStochastics1986} and Bandi and Moloche. \cite{BandiEconometTheor2018}  Kernel-based estimators have several advantages over their Bayesian counterparts. In addition to being simpler and easier to implement, their asymptotic errors are well understood. Finally, they can be readily applied in non-stationary systems wherein diffusivity and drift are also time-dependent,\cite{LamourouxPhysRevA2009} or when diffusion is anomalous.\cite{SaussereauBernoulli2014} It is indeed straightforward to relate isotropic diffusivity profiles to density fluctuations in a particulate system.\cite{EmbacherProcRoySocAMathPhys2018}

In these {\color{black}two} papers, we {\color{black}apply} the idea of kernel-based estimators to compute spatial profiles of the diffusivity tensor from single particle trajectories. The estimators that we derive and validate  are general enough to capture both spatial variations and directionality.  In {\color{black}the first} paper {\color{black}(i.e.,~the current one)}, we benchmark and validate this estimator using trajectories generated from SDEs according to known anisotropic diffusivity profiles. The next paper is dedicated to adapting these estimators to compute diffusivity profiles from molecular dynamics simulations.  

This paper is organized as follows. In Section~\ref{section:math-framework}, the mathematical derivation of the proposed estimator is presented, with details of numerical implementation and SDE integration outlined in Section~\ref{section:numerical-details}. In Section~\ref{section:results}, we validate the proposed estimator through a number of numerical tests, and assess its sensitivity to implementation details as well as the nature of confinement.  Section~\ref{section:conclusions} is dedicated to concluding remarks.

\section{Derivation of diffusivity estimators}
\label{section:math-framework}

 \noindent
 As mentioned above, the statistics of the spatiotemporal evolution of a typical fluid particle within a confined fluid can be described by Eq.~\eqref{Smoluch}. This implies that $\textbf{X}_t$, the trajectory of such a particle, has to be a Markov process that does not undergo discontinuous jumps. It can  be demonstrated that individual realizations of $\textbf{X}_t$ can be generated using the following SDE,\cite{ComerJCTC2014}
\begin{eqnarray}
d\textbf{X}_t&=&\left[-\beta \textbf{D}(\textbf{X}_t)\cdot\nabla U(\textbf{X}_t)+\nabla\cdot \textbf{D}(\textbf{X}_t)\right]dt\notag\\
&&+\sqrt{2\textbf{D}(\textbf{X}_t)}d\textbf{W}_t.\label{Overdamped Langevin}
\end{eqnarray}
with $\textbf{W}_t$ the Weiner process (i.e.,~the standard white noise). It must be noted that Eq.~(\ref{Overdamped Langevin}) is expressed in the It\^{o} convention,~i.e.,~that both drift and diffusivity are estimated at the beginning of the integration window. In general, the single-particle trajectories obtained from molecular simulations are statistically distinct from realizations of Eq.~(\ref{Overdamped Langevin}) at least at short timescales. As such, the task of developing accurate estimators of $\textbf{D}(\textbf{r})$ from such trajectories is a two-step process. The first step, which is the focus of the current paper, is aimed at validating and comparing different estimators in their ability to accurately compute $\textbf{D}(\textbf{r})$ from trajectories generated using Eq.~(\ref{Overdamped Langevin}). The second step, which will be the focus of the subsequent paper, will deal with subtleties of applying these estimators to trajectories that are qualitatively different from realizations of Eq.~(\ref{Overdamped Langevin}) at short observation windows.  

When it comes to the first question, we consider two different approaches. The first approach is based on the observation that over sufficiently short time windows, a particle will feel the local diffusivity at its starting point. As such, $\textbf{D}(\textbf{r})$ can be estimated from the second-order Kramers-Moyal coefficients as:
\begin{eqnarray}\label{jumpmoments}
\textbf{D}(\textbf{r}) &=& \lim_{h\rightarrow0^+}
\frac{\left\langle\left(\textbf{X}_{t+h}-\textbf{X}_t\right)\left(\textbf{X}_{t+h}-\textbf{X}_t\right)^\dagger \right\rangle_{\textbf{X}_t=\textbf{r}}}{2h}\label{jumpmoments}
\end{eqnarray} 
In theory, Eq.~(\ref{jumpmoments}) is sufficient for estimating $\textbf{D}(\textbf{r})$. In practice though, the  single-particle trajectories emanating from tracking experiments or molecular simulations will visit each position $\textbf{r}$  according to a probability distribution $p_0(\textbf{r})$ that is dictated by the relevant thermodynamic ensemble. Therefore, the pointwise expectation of Eq.~(\ref{jumpmoments}) will turn into the following integral equation:
\begin{eqnarray}
&&\int\textbf{D}(\textbf{r})p_0(\textbf{r})d^3\textbf{r}= \notag\\&&
\lim_{h\rightarrow0^+}
\frac{\left\langle\left(\textbf{X}_{t+h}-\textbf{X}_t\right)\left(\textbf{X}_{t+h}-\textbf{X}_t\right)^\dagger \right\rangle_{\textbf{X}_t\sim p_0(\textbf{r})}}{2h} \label{eq:intD}
\end{eqnarray}
One can, however, derive a localized form of (\ref{eq:intD}) by partitioning the observation domain into sufficiently small bins. More precisely, if the starting points of individual realizations of (\ref{Overdamped Langevin}) are drawn from $p_1(\textbf{r}):=p_0(\textbf{r})\chi_b(\textbf{r})/\int p_0(\textbf{s})\chi_b(\textbf{s})d\textbf{s}$ with $\chi_b(\textbf{r})$ the indicator function of bin $b$, Eq.~(\ref{eq:intD}) can  be rewritten as:
\begin{eqnarray}
&&\frac{\int_b\textbf{D}(\textbf{r})p_0(\textbf{r})d^3\textbf{r}}{\int_bp_0(\textbf{r})d^3\textbf{r}}= \notag\\&& \lim_{h\rightarrow0^+}
\frac{\left\langle\left(\textbf{X}_{t+h}-\textbf{X}_t\right)\left(\textbf{X}_{t+h}-\textbf{X}_t\right)^\dagger \right\rangle_{\textbf{X}_t\sim p_1(\textbf{r})}}{2h}\label{eq:int-D-confined}
\end{eqnarray} 
Note that the integral on the left hand side is the average diffusivity within $b$ and converges to pointwise diffusivity when $\int p_0(\textbf{r})\chi_b(\textbf{r})d^3\textbf{r}\rightarrow0$. Therefore, for sufficiently small bins and sufficiently short observation windows, the following estimator can be constructed for $\widehat{\textbf{D}}_b$, the pointwise diffusivity within bin $b$:
\begin{eqnarray}
\widehat{\textbf{D}}_b &=& \frac{1}{2h}
\frac{\sum_{i=1}^N\left(\textbf{X}_{i,t+h}-\textbf{X}_{i,t}\right)\left(\textbf{X}_{i,t+h}-\textbf{X}_{i,t}\right)^{\dagger}\chi_b\left(\textbf{X}_{i,t}\right)}
{\sum_{i=1}^N\chi_b\left(\textbf{X}_{i,t}\right)}\notag\\
&&\label{eq:LCE-estimator}
\end{eqnarray}
with $N$ the number of trajectories generated using (\ref{Overdamped Langevin}). It must be noted that both Eqs.~(\ref{eq:int-D-confined}) and (\ref{eq:LCE-estimator}) can be viewed as rigorous generalizations of the notion of an \emph{ad hoc} local mean-square displacement. The estimator in (\ref{eq:LCE-estimator}) relates local diffusivity to a covariance matrix constructed from short-time displacements and is therefore referred to as a \emph{local covariance estimator (LCE)}. 

\begin{algorithm*}
	\caption{\label{alg:fce}Algorithmic details of implementing the FCE estimator given by Eq.~(\ref{eq:FCE-estimator}).} 
   \begin{algorithmic}[1]
   	\State\textbf{Procedure} EstimateDiffusivity ($\{\textbf{X}_{i,t}\}_{i=1}^N$, $\textbf{r}_0$, $\epsilon$, $h$, $\Delta t$) \Comment{$\textbf{X}_i\equiv (\textbf{X}_{i,0},\cdots,\textbf{X}_{i,n_t})$'s are discrete-time trajectories with both {\color{black}$n_t=t/\Delta t$} and {\color{black}$n_h=h/\Delta{t}$} integers.}
	\State Choose $S=\{\textbf{k}_i\}_{i=1}^m$, a stencil of unit vectors for which $\mathcal{K}$ given in Eq.~(\ref{eq:mathcal-K}) is a full-rank matrix.
	\State Choose $g(\mathbf{r})\geq 0$ to be a $C^1$ function such that $g(\pmb0)>0$ and $\int g(\mathbf{r})\,d^3\mathbf{r}=1$  
	\State Construct an $\textbf{r}_0$-centered and $\epsilon$-dilated version of $g(\textbf{r})$,~i.e.,~$G_{\epsilon{\color{black}, \mathbf{r}_0}}(\textbf{r}) = \epsilon^{-d}g[(\textbf{r}-\textbf{r}_0)/\epsilon]$. 
	\State $\rho:=0$. \Comment{$\rho$ is the denominator of Eq.~(\ref{eq:FCE-estimator})}.
	\For{$i=1,2,\cdots,N$}
		\For{$\tau=0,1,\cdots,{\color{black}
n_t}$}
			\State $\rho:=\rho + G_{\epsilon{\color{black}, \mathbf{r}_0}}(\textbf{X}_{i,\tau})$.
		\EndFor
	\EndFor
	\State $\rho:=\rho/\left[N({\color{black}n_t}+1)\right]$.
	\For{$j=1,2,\cdots,m$}
		\State $\mathcal{P}_{j}:=0$.
		\State $\gamma_j(\textbf{r}):=f_{\textbf{k}_j}(\textbf{r})G_{\epsilon{\color{black}, \mathbf{r}_0}}(\textbf{r})$.
		\For{$i=1,2,\cdots,N$}
			\For{$\tau=0,1,\cdots,{\color{black}n_t-n_h}$}
				\State $\mathcal{P}_j:=\mathcal{P}_j+\Delta_h\gamma_j^*(\textbf{X}_{i,\tau}) \Delta_hf_{\textbf{k}_j}(\textbf{X}_{i,\tau})$.
			\EndFor
		\EndFor
		\State $\mathcal{P}_j:=\mathcal{P}_j/\left[N({\color{black}n_t-n_h})\right]$.
		\State $\mathcal{P}_j:=\Re\left[\mathcal{P}_j\right]/\left[2\alpha^2h\right]$.
	\EndFor
	\State Solve the linear system $\mathcal{K}\mathcal{D}=\mathcal{P}$ exactly (for $m=6$) or using a least-square approach (for $m>6$).
	\State\Return $\mathcal{D}$.
   \end{algorithmic}
\end{algorithm*}

An alternative approach, which is our method of choice in this work, involves using a \emph{filter function} to capture particle displacements in select regions and along select directions, and leads to a family of estimators known as \emph{filtered covariance estimators (FCEs)}. More precisely, for  a piecewise $C^2$ filter function $\gamma:\R^3\rightarrow \mathbb{C}$,  ${\color{black}Y_t:=}\gamma(\textbf{X}_t)$ will {\color{black}evolve according to the following stochastic process given by} It\^{o}'s Lemma,\cite{SchussTheoryStochastic2010}
\begin{eqnarray}
{\color{black}dY_t} 
&{\color{black}=}&
 {\color{black}\left[\nabla\gamma(\mathbf{X}_t)\cdot q(\mathbf{X}_t)+\mathbf{D}(\mathbf{X}_t):\mathbf{H}_\gamma(\mathbf{X}_t)\right] dt 
 }\notag\\
 && {\color{black}+\nabla\gamma(\mathbf{X}_t)\cdot\sqrt{2\mathbf{D}(\mathbf{X}_t)}dW_t}
 \label{eq:GammaX-SDE}
\end{eqnarray}
{\color{black}where $q(\textbf{r})=-\beta\mathbf{D}\cdot\nabla U+\nabla\cdot\mathbf{D}$ and $\mathbf{H}_\gamma$ is the Hessian of the filter function $\gamma(\cdot)$.}
The short-time evolution of $\gamma(\textbf{X}_t)$ can {\color{black}therefore} be related to the 'filtered` diffusivity as follows:\cite{SchussTheoryStochastic2010}
\begin{eqnarray}
&&\int \nabla\gamma^\dagger(\textbf{r})\textbf{D}(\textbf{r})\nabla\gamma(\textbf{r}){\color{black}p_0(\textbf{r})}\,d^3\textbf{r} =  \notag\\
&& \lim_{h\rightarrow0^+} \frac{\left\langle\left|\gamma(\textbf{X}_{t+h})-\gamma(\textbf{X}_{t})\right|^2 \right\rangle_{\textbf{X}_t\sim p_0(\textbf{r})}}{2h}\label{AverGammaC}
\end{eqnarray}
It must be noted that since the filter function maps a vectorial trajectory onto a scalar quantity, a single filter will no longer be sufficient for resolving the anisotropy of the diffusivity tensor, but a sufficient number of filters (each capturing a different direction) can provide an accurate estimate of $\textbf{D}(\textbf{r})$.  Moreover, by spatially filtering for particles that are sufficiently close to the measurement point both in the beginning and the end of the observation window $h$, FCEs generally have a smaller localization error than the LCE given by Eq.~(\ref{eq:LCE-estimator}) {\color{black}as will be demonstrated in Fig.~\ref{eps_eff_1} and Appendix~\ref{appendix2}}.

\begin{algorithm*}
{\color{black}
	\caption{\label{alg:fce-alpha0}{\color{black}Algorithmic details of implementing the FCE estimator given by Eq.~(\ref{eq:FCE-estimator-alpha0}).}} 
   \begin{algorithmic}[1]
   	\State\textbf{Procedure} EstimateDiffusivity ($\{\textbf{X}_{i,t}\}_{i=1}^N$, $\textbf{r}_0$, $\epsilon$, $h$, $\Delta t$) \Comment{$\textbf{X}_i\equiv (\textbf{X}_{i,0},\cdots,\textbf{X}_{i,n_t})$'s are discrete-time trajectories with both $n_t=t/\Delta{t}$ and $n_h=h/\Delta{t}$ integers.}
	\State Choose $S=\{\textbf{k}_i\}_{i=1}^m$, a stencil of unit vectors for which $\mathcal{K}$ given in Eq.~(\ref{eq:mathcal-K}) is a full-rank matrix.
	\State Choose $g(\mathbf{r})\geq 0$ to be a piecewise $C^1$ function such that $g(\pmb0)>0$ and $\int g(\mathbf{r})\,d^3\mathbf{r}=1$  
	\State Construct an $\textbf{r}_0$-centered and $\epsilon$-dilated version of $g(\textbf{r})$,~i.e.,~$G_{\epsilon{\color{black}, \mathbf{r}_0}}(\textbf{r}) = \epsilon^{-d}g[(\textbf{r}-\textbf{r}_0)/\epsilon]$. 
	\State $\rho:=0$. \Comment{$\rho$ is the denominator of Eq.~(\ref{eq:FCE-estimator-alpha0})}.
	\For{$i=1,2,\cdots,N$}
		\For{$\tau=0,1,\cdots,n_t$}
			\State $\rho:=\rho + G_{\epsilon{\color{black}, \mathbf{r}_0}}(\textbf{X}_{i,\tau})$.
		\EndFor
	\EndFor
	\State $\rho:=\rho/\left[N(n_t+1)\right]$.
	\For{$j=1,2,\cdots,m$}
		\State $\mathcal{P}_{j}:=0$.
		\For{$i=1,2,\cdots,N$}
			\For{$\tau=0,1,\cdots,n_t-n_h$}
				\State $\mathcal{P}_j:=\mathcal{P}_j+\left[
				G_{\epsilon{\color{black}, \mathbf{r}_0}}(\textbf{X}_{i,\tau+n_h})+G_{\epsilon{\color{black}, \mathbf{r}_0}}(\textbf{X}_{i,\tau})
				\right]\left[\textbf{k}\cdot\Delta_h\textbf{X}_{i,\tau}\right]^2$.
			\EndFor
		\EndFor
		\State $\mathcal{P}_j:=\mathcal{P}_j/\left[4hN(n_t-n_h)\right]$.
	\EndFor
	\State Solve the linear system $\mathcal{K}\mathcal{D}=\mathcal{P}$ exactly (for $m=6$) or using a least-square approach (for $m>6$).
	\State\Return $\mathcal{D}$.
   \end{algorithmic}
   }
\end{algorithm*}

The filter functions that we employ in this work are of the form $\gamma_{\textbf{k}}(\textbf{r}) := f_{\textbf{k}}(\textbf{r})G(\textbf{r})$. Here, $f_{\textbf{k}}(\textbf{r}):=e^{-i\alpha\textbf{k}\cdot\textbf{r}}$ gives the filter function its directionality wherein $\textbf{k}\in\mathbb{R}^3$ is a fixed unit vector. $G(\textbf{r})$, however, is a localized real-valued function that selects for the particles that reside within a particular region of the observation domain. {\color{black}In order to obtain pointwise estimates of diffusivity, we choose} $G(\textbf{r}) := g(\textbf{r}-\textbf{r}_0)$ {\color{black}so that} $g(\textbf{r})$ is a real-valued bounded function with a compact support and $\textbf{r}_0$ is the  point around which local diffusivity is estimated. {\color{black}It is generally} more convenient for $g(\textbf{r})$ to approximate the Dirac delta function. {\color{black}As we will discuss later, however,}  $g(\cdot)$ {\color{black}does not need to have compact support and can instead}  belong to a set of orthogonal basis functions. Moreover, it is more convenient to choose a particular  functional form for $g(\textbf{r})$ and then define $G(\textbf{r})$ as an $\epsilon$-dilated version of it, namely  $G_\epsilon(\textbf{r}) = \epsilon^{-d}g\left[(\textbf{r}-\textbf{r}_0)/\epsilon\right]$. That way, the width of the kernel can be readily adjusted by altering $\epsilon$. For this particular choice of $\gamma_{\textbf{k}}(\textbf{r})$, Eq.~(\ref{AverGammaC}) can be rearranged to conclude that (Appendix~\ref{appendix1}):
\begin{eqnarray}
&& \int \textbf{k}^T\textbf{D}(\textbf{r})\textbf{k}\,p_0(\textbf{r})G(\textbf{r}) \,d^3\textbf{r} = \notag\\
&& \lim_{h\rightarrow0^+} \frac{\left\langle\left[\gamma_{\textbf{k}}(\textbf{X}_{t+h})-\gamma_{\textbf{k}}(\textbf{X}_{t})\right]^*\left[f_{\textbf{k}}(\textbf{X}_{t+h})-f_{\textbf{k}}(\textbf{X}_{t})\right] \right\rangle^{\Re}_{p_0}}{2\alpha^2 h}\notag\\
&& \label{AverGammaC2}
\end{eqnarray}
where $\langle\rangle^{\Re}_{p_0}$ corresponds to the real part of a complex-valued expectation estimated for $\textbf{X}_t\sim p_0(\cdot)$. The validity of (\ref{AverGammaC2}) at the limit of $h\rightarrow0^+$ implies that the following approximation can be made for a sufficiently small $h$:
\begin{eqnarray}
&& \frac{\left\langle\Delta_h\gamma_{\textbf{k}}^*(\textbf{X}_t)\Delta_hf_{\textbf{k}}(\textbf{X}_t)\right\rangle^{\Re}_{\textbf{X}_t\sim p_0(\textbf{r})}}{2\alpha^2h\langle G(\textbf{X}_t)\rangle_{\textbf{X}_t\sim p_0(\textbf{r})}} \approx \notag\\
&& \frac{\bigintss \textbf{k}^T\textbf{D}(\textbf{r})\textbf{k}\,G(\textbf{r})p_0(\textbf{r})\,d^3\textbf{r}}{\bigintsss G(\textbf{r})p_0(\textbf{r})\,d^3\textbf{r}}\label{Localized_gammaC}
\end{eqnarray}
where $\Delta_hF(\textbf{X}_t):=F(\textbf{X}_{t+h})-F(\textbf{X}_t)$. Note that the righthand side of Eq.~(\ref{Localized_gammaC}) is the weighted average of $D_{\textbf{k}\textbf{k}}=\textbf{k}^T\textbf{D}(\textbf{r})\textbf{k}$ within the region probed by $G(\textbf{r})$, which will converge to the pointwise diffusivity along $\textbf{k}$ upon letting $\epsilon\rightarrow0$. An FCE estimator of $D_{\textbf{kk}}$ can therefore be formulated as:
\begin{eqnarray}
\widehat{D}_{\textbf{kk}} &=& \frac1{2\alpha^2h}
\frac{\Re\left[\sum_{i=1}^N\Delta_h\gamma_{\textbf{k}}^*(\textbf{X}_{i,t})\Delta_hf_{\textbf{k}}(\textbf{X}_{i,t})\right]}{\sum_{i=1}^NG(\textbf{X}_{i,t})}\label{eq:FCE-estimator}
\end{eqnarray}
{\color{black}
One can, however, derive another estimator by observing the limiting behaviors of (\ref{AverGammaC2}) and (\ref{eq:FCE-estimator}) at $\alpha\rightarrow0$ (Appendix~\ref{appendix:alpha0}):
\begin{eqnarray}
&& \int \textbf{k}^T\textbf{D}(\textbf{r})\textbf{k}\,p_0(\textbf{r})G(\textbf{r}) \,d^3\textbf{r} = \notag\\
&& \lim_{h\rightarrow0^+} \frac{\left\langle\left[ 
G(\textbf{X}_{t+h})+G(\textbf{X}_t)
\right]\left[\mathbf{k}\cdot\Delta_h\mathbf{X}_t\right]^2 \right\rangle_{p_0}}{4h}\notag\\
&& \label{AverGammaC2-alpha0}
\end{eqnarray}
which, in the case of a localized kernel, yields the  $\alpha\rightarrow0$ equivalent of (\ref{eq:FCE-estimator}):
\begin{eqnarray}
\widehat{D}_{\textbf{kk}}^{\alpha\rightarrow0} &=& \frac1{4h}
\frac{\sum_{i=1}^N\left[G(\textbf{X}_{i,t+h})+G(\textbf{X}_{i,t})\right]
\left[
\textbf{k}\cdot\Delta_h\textbf{X}_{i,t}
\right]^2}{\sum_{i=1}^NG(\textbf{X}_{i,t})}
\notag\\
&& \label{eq:FCE-estimator-alpha0}
\end{eqnarray}
It must be noted that Eq.~(\ref{eq:FCE-estimator-alpha0}) is remarkably simple, and states that local diffusivity along a particular direction $\textbf{k}$ is the weighted average of mean-squared displacement projected along $\textbf{k}$ wherein particles that are present at a particular point either at the beginning or at the end of an observation window contribute equally to MSD. In essence, Eq.~(\ref{eq:FCE-estimator-alpha0})  provides a natural-- and yet rigorous-- means of constructing an \emph{ad hoc} local MSD. One could conceivably postulate Eq.~(\ref{eq:FCE-estimator-alpha0}) based solely on heuristics, but our procedure of deriving it starting from the complex valued estimator based on $f_\mathbf{k}(\mathbf{x})=e^{-i \alpha\mathbf{k}\cdot\mathbf{x}}$ can be viewed as a more rigorous justification for it.}

In order to reconstruct the full diffusivity tensor at $\textbf{r}_0$ {\color{black}for either estimator}, one needs  to choose $S=\{\mathbf{k}_i\}_{i=1}^m$, a stencil of unit vectors, and estimate $\widehat{D}_{\textbf{k}_i\textbf{k}_i}$ for each $\mathbf{k}_i\in S$. According to the Onsager reciprocity principle,\cite{OnsagerPhysRev1931} the diffusivity tensor needs to be symmetric and can have a maximum of six independent components in three dimensions.  As such, a minimum of six stencil vectors will be needed  and the task of reconstructing $\widehat{\textbf{D}}(\textbf{r}_0)$ will therefore involve solving the linear system $\mathcal{K}\mathcal{D} = \mathcal{P}$ where $\mathcal{K}$ is an $m\times 6$ matrix with following components:
\begin{subequations}\label{eq:mathcal-K}
\begin{eqnarray}
\mathcal{K}_{i,1} &=& k_{i,x}^2 \\
\mathcal{K}_{i,2} &=& k_{i,y}^2 \\ 
\mathcal{K}_{i,3} &=& k_{i,z}^2 \\ 
\mathcal{K}_{i,4} &=& 2k_{i,x}k_{i,y} \\ 
\mathcal{K}_{i,5} &=& 2k_{i,x}k_{i,z} \\ 
\mathcal{K}_{i,6} &=& 2k_{i,y}k_{i,z} 
\end{eqnarray}
\end{subequations}
and $\mathcal{D}$ is the following column vector:
\begin{eqnarray}
\mathcal{D}&=&\left[
\begin{array}{c}
\widehat{D}_{xx}\\
\widehat{D}_{yy}\\
\widehat{D}_{zz}\\
\widehat{D}_{xy}\\
\widehat{D}_{xz}\\
\widehat{D}_{yz}\\
\end{array}
\right] .
\end{eqnarray}
$\mathcal{P}$, however, is an $m\times 1$ matrix whose entries are the right hand side of (\ref{eq:FCE-estimator}) {\color{black}or (\ref{eq:FCE-estimator-alpha0})} for each filter function. As long as $\mathcal{K}$ is a full-rank matrix, the associated linear system can be solved exactly (in the case of $m=6$) or using a least-square approach (if $m>6$). For systems that are stationary, both LCE and FCE estimators can also be averaged over time in order to obtain more stable estimates.  {\color{black}Algorithms~\ref{alg:fce} and ~\ref{alg:fce-alpha0} describe procedures for obtaining FCE estimates of $\mathcal{D}$ using the estimators given by Eqs.~(\ref{eq:FCE-estimator}) and (\ref{eq:FCE-estimator-alpha0}) with time averaging, respectively.}

\section{Numerical validation of diffusivity estimators}
\label{section:numerical-details}


\noindent
In order to asses the performance of the proposed estimators, we perform numerical experiments in which single-particle trajectories are generated using the SDE given by Eq.~(\ref{Overdamped Langevin}) according to known anisotropic and position-dependent diffusivity tensors. Each estimator is then assessed based on its ability to accurately back-compute the  profile utilized for generating the trajectories.  We use the open-source \textsc{Julia} package, \texttt{DifferentialEquations.jl},\cite{RackauckasJOpenResSoft2017} for numerically integrating the SDEs using an explicit high-order Runge-Kutta discretization and with a time step of $\Delta{t}=10^{-6}$. Unless otherwise specified, a total of $N=14,400$ trajectories were generated per profile, each integrated for a minimum of 5,000 time steps. Despite their different mathematical forms, the considered diffusivity profiles are all anisotropic, diagonal and functions of $z$ only.  The integration domain is periodic along the $z$ direction. As such, all diffusivity profiles are periodized using the following procedure. For a non-periodic profile, $\textbf{D}(z)$ defined over the domain  $z\in [0,a]$, its periodic extension $\textbf{D}_p(z)$ is defined as:
\begin{eqnarray}
\textbf{D}_{\text{p}} (z) &=& \left\{
\begin{array}{lll}
\textbf{D}(0) && |z| < b \\
\textbf{D}(z-b) && b\le z\le a+b\\
\textbf{D}(-z-b) && -a-b\le z\le -b\\
\textbf{D}(a) && a+b<|z|\le a+2b\\
\end{array}
\right.
\end{eqnarray}
Here, $b=2$ is the width of a narrow buffer region that is added between each profile and its periodic image. {\color{black}(We choose a value of $b=0$ if the original diffusivity profile is periodic.)} During integration, particles that leave the simulation box along the $z$ direction are folded back into the domain from the other side. Unless otherwise specified, all trajectories are generated under zero drift,~i.e.,~$U(z)=0$.

\begin{table}
	\centering
	\caption{Default implementation parameters for the utilized FCE estimator. \label{table:defaults}}
	\begin{tabular}{ll|ll}
	\hline\hline
	Parameter~~~& Value~~~ & Parameter~~~ & Value~~~  \\
	\hline
	$\epsilon$ & $0.2637$ & $h$ & $10^{-4}$  \\
	$\alpha$ & $0$ & $N$ & $14,\!400$\\
	$m$ & 13 & $g(z)$ & $(1-|z|)H(1-|z|)$\\
	\hline
	\end{tabular}
\end{table}

In estimating the diffusivity profile from the generated trajectories, we use an observation window of $h=10^{-4}$ (i.e.,~100 time steps) unless otherwise specified. In the case of the FCE estimator of Algorithms~\ref{alg:fce} {\color{black} and~\ref{alg:fce-alpha0}}, we use stencils of different sizes and compositions, either generated randomly (by generating $m\ge6$ vectors such that each component is sampled from a standard normal distribution) or via a deterministic approach described below. For the majority of the calculations conducted here, we use a deterministic stencil of $m=13$ unit vectors obtained as follows. First, we start with the 26 vectors at the edges, vertices and faces of a unit cube, and eliminate one out of each pair related to one another via inversion. We also consider deterministic stencils of 6 and 10 vectors by choosing half the vectors connecting the geometric center of a dodecahedron and icosahedron to the centers of its faces, respectively.  All default implementation details for the FCE estimator are given in Table~\ref{table:defaults}.

{\color{black}
To estimate the error bars in the current study, we partition the $N$ trajectories obtained from SDE integration into five blocks of equal sizes and subsequently apply the corresponding estimator to each block. The resulting profiles are utilized to compute the mean and standard error of the data. Notably, it should be emphasized that all the reported error bars  correspond to 95\% confidence intervals, which are equivalent to twice the calculated standard error.
}

\begin{figure}
	\centering
	\includegraphics[width=0.5\textwidth]{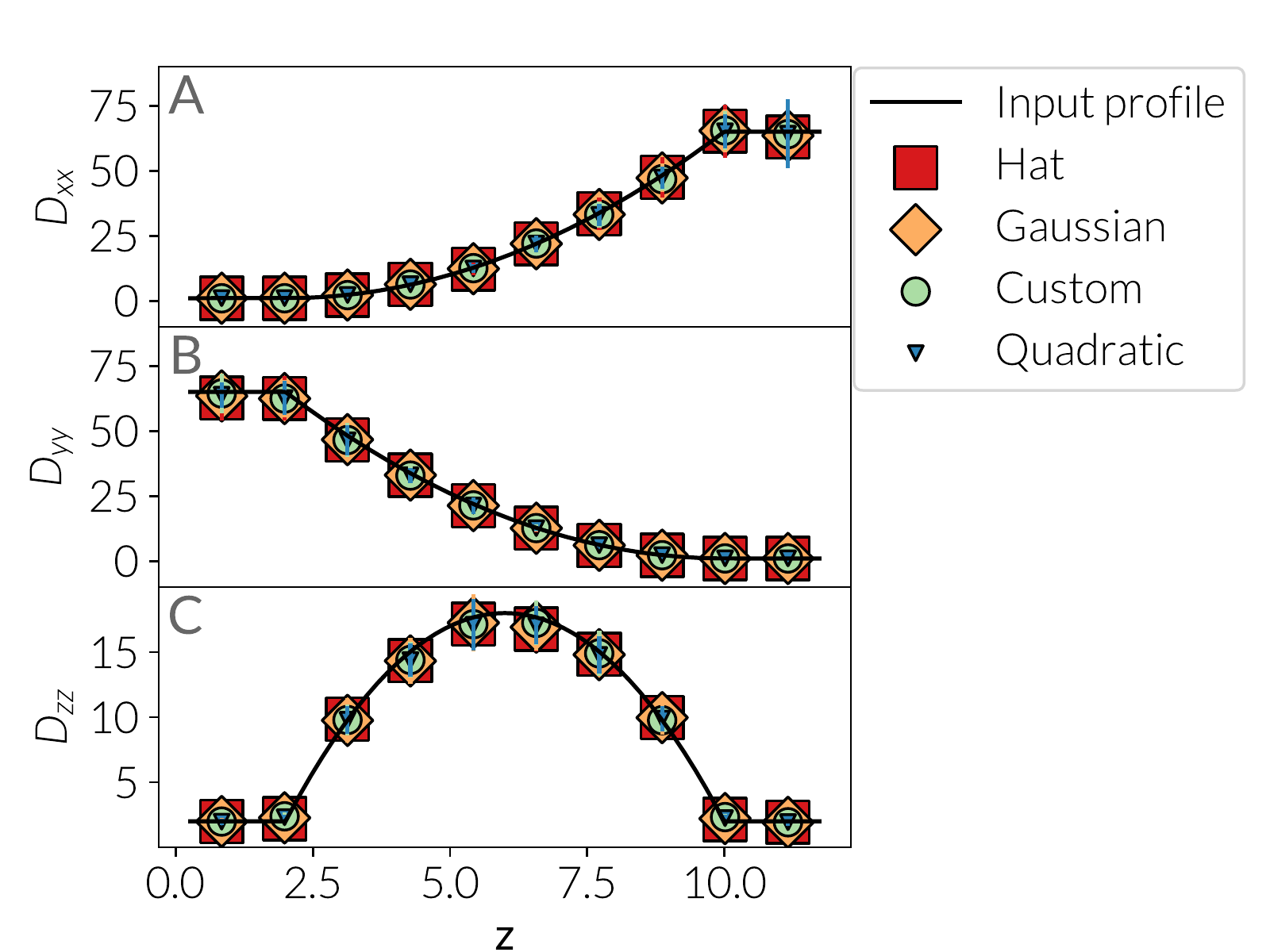}
	\vspace{-15pt}
	\caption{The effect of the localization function, $g(\cdot)$, on the accuracy of the employed FCE in estimating the (A) $D_{xx}$, (B) $D_{yy}$ and (C) $D_{zz}$ components of an anisotropic quadratic diffusivity profile. {\color{black}Here, 'Hat` and 'Custom` refer to the kernels (a) and (c) given at the start of Section~\ref{section:implementation-details}.}}
	\label{fig:diffusivitykerneltype_1}
	\vspace{-15pt}
\end{figure}

\begin{figure*}
	\centering
	\includegraphics[width=0.95\textwidth]{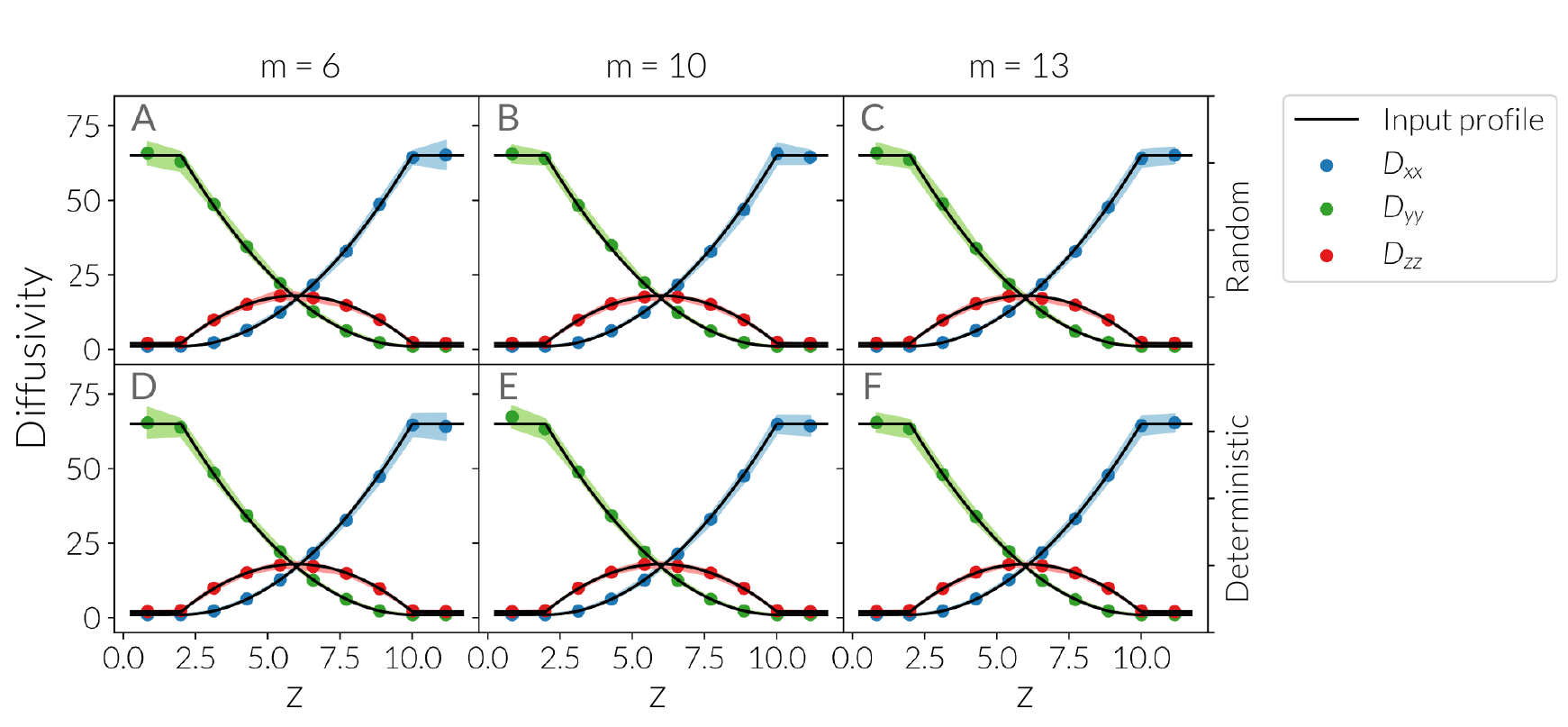}
	\caption{Sensitivity of the FCE estimator to the size and composition of the stencil. (A-C) Random and (D-F) deterministic stencils with (A, D) $m=6$, (B, E) $m=10$ and (C, F) $m=13$ vectors are considered.}
	\label{fig:stencil-size-composition}
\end{figure*}

\section{Results and discussions}
\label{section:results}

{\color{black}
\noindent 
Due to its numerical stability and simplicity, we primarily validate the $\alpha\rightarrow0$ estimator given by Eq.~(\ref{eq:FCE-estimator-alpha0}). However, in certain situations, the $\alpha\neq0$ estimator from Eq.~(\ref{eq:FCE-estimator}) may offer advantages by eliminating spurious outliers that could bias the diffusivity estimate in the absence of the complex exponential.Therefore, we also evaluate the numerical stability and sensitivity of the latter estimator. It should be noted, however, that in all the validations presented here, the performances of the two estimators are nearly indistinguishable.

}

\subsection{Sensitivity to implementation details}
\label{section:implementation-details}
\noindent
We first explore the sensitivity of the FCE estimator {\color{black}given in Eq.~(\ref{eq:FCE-estimator-alpha0})} to properties of the filter function. Our first test is to probe the effect of the localization function $g(\cdot)$ on the accuracy of the estimate. As is generally the case with  Nadaraya-Watson type estimators, we do not expect the diffusivity estimate to be strongly impacted by  the particular choice of the kernel function. Indeed, we demonstrate in Appendix~\ref{appendix2} that the systematic error introduced due to the particular form of the localization function scales as $O(\epsilon^2)$. As can be noted in Eq.~(\ref{eq:localization-error-analysis}),  the prefactor to the leading $\epsilon^2$ term only depends on the second moment of the localization function given by Eq.~(\ref{eq:g-2nd-mom}).  Therefore, if $\epsilon$ is sufficiently small, the choice of the kernel will not have a significant impact on the estimate. We demonstrate this numerically by probing SDE trajectories generated according to a quadratic diffusivity profile using the following four localization functions:
\begin{itemize}
	\item [(a)] $g(z)=(1-|z|)H(1-|z|)$
	\item [(b)] $g(z) = \frac{1}{\sqrt{2\pi}}e^{-z^2/2}$
	\item [(c)] $g(z)=e^{-\frac{1}{(1-|z|^2)^2}}H(1-|z|)$
	\item [(d)] $g(z)=(1-z^2)H(1-|z|)$, 
\end{itemize}
where $H(z)$ is the Heaviside step function. As can be seen in Fig.~\ref{fig:diffusivitykerneltype_1}, the diffusivity estimates obtained from these four kernels are virtually indistinguishable. This implies that the choice of the kernel function can be merely made based on convenience and ease of implementation as long as it satisfies the properties outlined in Appendix~\ref{appendix1}. As such, we use (a) as our kernel in the remainder of this work.

\begin{figure}
	\centering
	\includegraphics[width=0.46\textwidth]{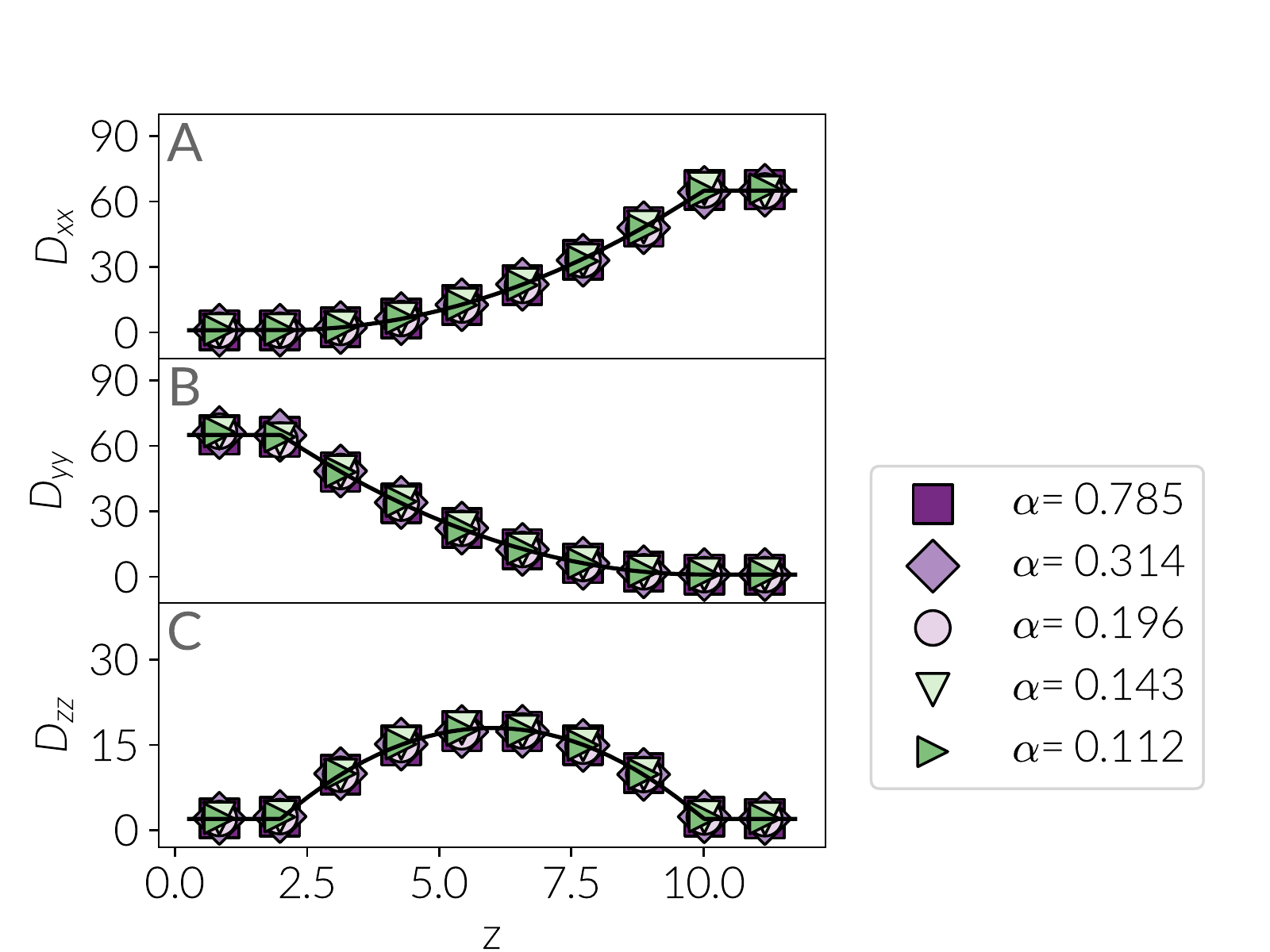}
	\caption{The effect of $\alpha$ on the accuracy of the employed FCE in estimating the (A) $D_{xx}$, (B) $D_{yy}$ and (C) $D_{zz}$ components of an anisotropic quadratic diffusivity profile using a deterministic stencil of $m=13$ vectors. } 
	\label{fig:alpha-effect}
	\vspace{-10pt}
\end{figure}

\begin{figure*}
	\centering
	\includegraphics[width=.7\textwidth]{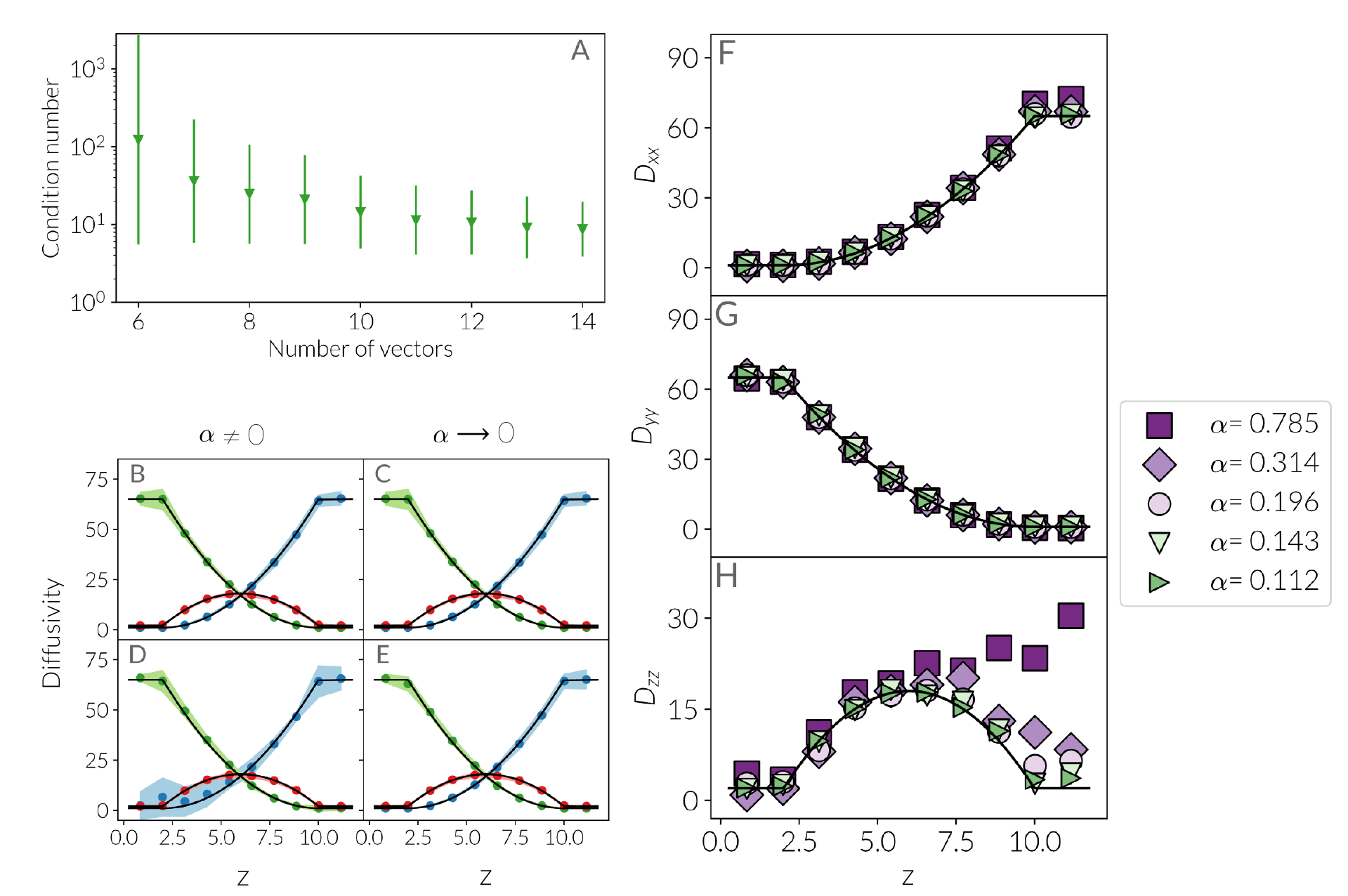}
	\caption{The effect of $\kappa(\mathcal{K})$, the condition number of the matrix $\mathcal{K}$, on the performance of the FCE estimator. (A) $\langle\kappa(\mathcal{K})\rangle$ vs.~$m$, the number of random vectors in the stencil, obtained from 100 independent draws for each $m$. (B-E) Diffusivity estimates obtained from two random stencils of $m=6$ vectors with condition numbers  (B-C) 21 and (D-E) 2003 using (B,D) Eq.~(\ref{eq:FCE-estimator}) and (C,E) Eq.~(\ref{eq:FCE-estimator-alpha0}). (F-H) Sensitivity of FCE estimates of (F) $D_{xx}$, (G) $D_{yy}$ and (H) $D_{zz}$ to $\alpha$ upon using a random stencil with a large condition number. Using a large $\alpha$ results in an overestimation of $D_{zz}$ by an order of magnitude. 
 \label{fig:condition-number}}
\end{figure*}

We then evaluate the effect of stencil size and composition on the performance of the estimator. As can be seen in Fig.~\ref{fig:stencil-size-composition}, the FCE estimator performs well, irrespective of whether they are chosen {\color{black}randomly} (Figs.~\ref{fig:stencil-size-composition}A-C) or {\color{black}deterministically} (Figs.~\ref{fig:stencil-size-composition}D-F). This demonstrates that it is not necessary to identify stencils that contain special vectors, such as the eigenvectors of the diffusivity tensor. Therefore, the FCE estimator proposed here is expected to work in more complex geometries in which special directions such as diffusivity eigenvectors might also be position-dependent and therefore unknown.

We finally explore the effect of $\alpha$ on the performance of the estimator {\color{black}given by Eq.~(\ref{eq:FCE-estimator}). As outlined earlier, using such an estimator might offer advantages by mitigating the contributions of spurious outliers.} The parameter $\alpha$ determines the extent to which $f_\textbf{k}(\textbf{r})$ oscillates within the observation domain. Indeed, if $\alpha$ is too large, aliasing artifacts might arise due to particles moving over length scales comparable to the wavelength of $f_\textbf{k}(\textbf{r})$ during the observation window. Fig.~\ref{fig:alpha-effect} depicts how changing $\alpha$ affects the performance of the FCE estimator in reconstructing an anisotropic quadratic profile. {\color{black}Clearly, such aliasing effects are not noticeable, and the estimator given by Eq.~(\ref{eq:FCE-estimator}) performs as well as the $\alpha\rightarrow0$ estimator.

The main difference between these two estimators, however, is in their numerical stability, particularly when using stencils that contain fewer randomly chosen vectors. This is because the numerical stability of solving the exact matrix inversion problem (for $m=6$) and the least square optimization problem (for $m>6$) depends on the condition number of matrix $\mathcal{K}$ (defined as $\kappa(\mathcal{K})=\sigma_1/\sigma_6$ where $\sigma_i$ is the $i$-th largest singular value of $\mathcal{K}$). Fig.~\ref{fig:condition-number}A depicts, $\langle\kappa(\mathcal{K})\rangle_m$ the average condition number of  $\mathcal{K}$ constructed from a random stencil as a function of $m$. $\langle\kappa(\mathcal{K})\rangle_m$ is pathologically large for $m=6$, but decreases as more vectors are included. Yet, even for $m$'s as large as $8$, $\kappa(\mathcal{K})$ can still surpass $10^2$, which is generally considered as a soft threshold for assuring the numerical stability of solving linear systems. The effect of the condition number on the performance of the FCE estimator given by (\ref{eq:FCE-estimator}) can be seen in Figs.~\ref{fig:condition-number}B,D wherein diffusivity estimates have been obtained using a stencil of six random vectors but with different condition numbers. Clearly, the uncertainties are considerably larger for the $\mathcal{K}$ with the larger condition number (Fig.~\ref{fig:condition-number}D). Such susceptibilities are not discernible for deterministic stencils that generally have considerably smaller condition numbers. The three deterministic stencils employed in this work, in particular, have $\kappa(\mathcal{K})\approx2$. Such instabilities are barely noticeable for the $\alpha\rightarrow0$ estimator of (\ref{eq:FCE-estimator-alpha0}) as can be noted in Fig.~\ref{fig:condition-number}E. Nonetheless, introducing redundancy by including more vectors in the stencil can make the estimator given by (\ref{eq:FCE-estimator}) more robust to such instabilities.

Condition number of the stencil also impacts aliasing. As can be seen in Figs.~\ref{fig:condition-number}F-H, for instance, using too large of an $\alpha$ can lead to an overestimation of diffusivity by an order of magnitude when a poorly conditioned random stencil is uitlized. Moreover, such aliasing artifacts tend to impact the diffusivity prediction in an anisotropic manner. In Figs.~\ref{fig:condition-number}F-H, for instance, it is the $D_{zz}$ estimate that is most adversely impacted while aliasing artifacts are virtually nonexistent in $D_{xx}$ and $D_{yy}$ estimates. We must, however, note that this observation is due to the idiosyncrasy of the particular stencil employed here, and altering the composition of the stencil will also change the particular component(s) of the diffusivity tensor that will be impacted by large  $\kappa(\mathcal{K})$ and $\alpha$. 
 Using a sufficiently small $\alpha$, however, mitigates all these artifacts even for poor draws of random vectors. 
 }

\begin{figure}
	\centering
	\includegraphics[width=0.5\textwidth]{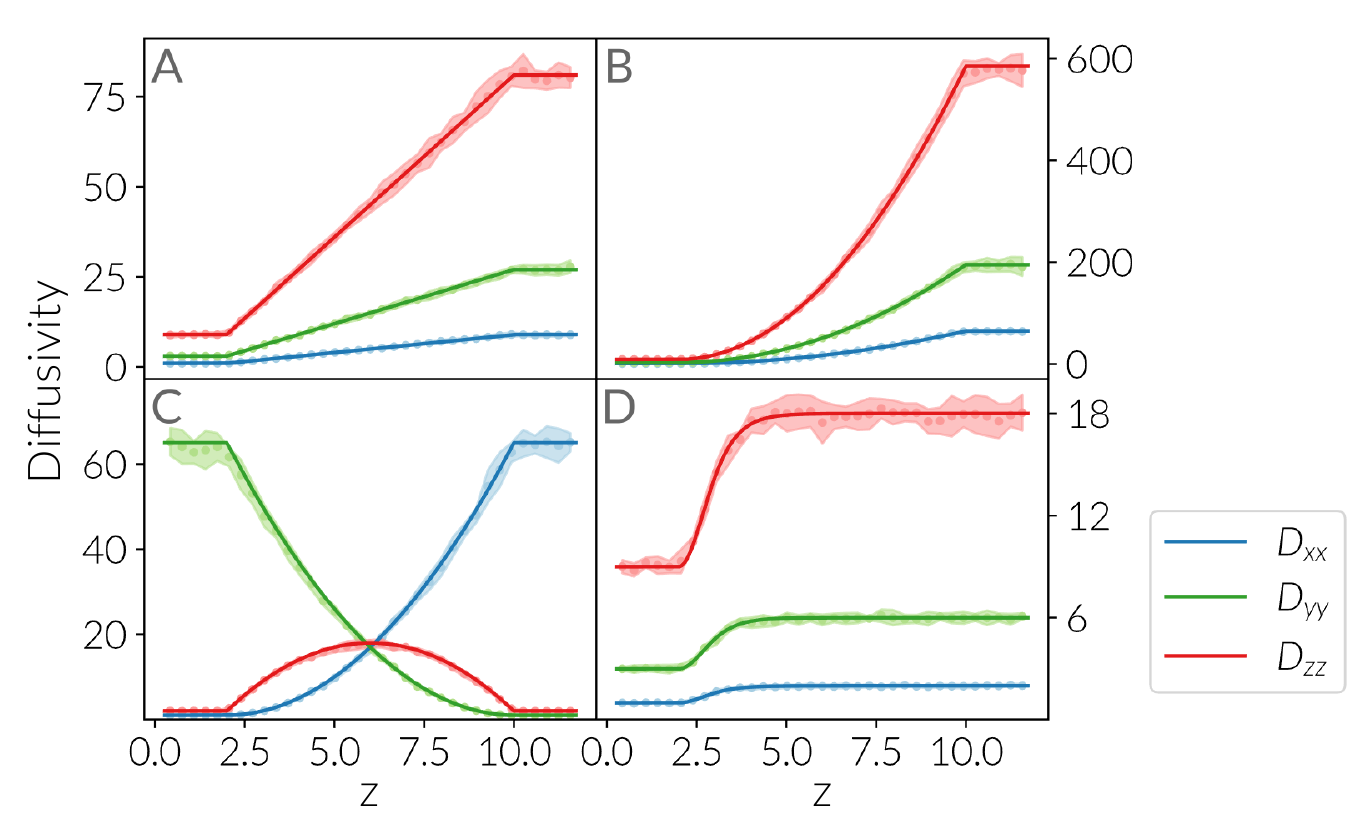}
	\caption{Diffusivity estimates obtained from trajectories generated in accordance to (A) linear, (B-C) quadratic and (D) tangent hyperbolic diffusivity profiles. The original profiles are depicted in solid lines while diffusivity estimates and their 95\% confidence intervals are depicted using light-colored circles and shades, respectively..}
	\label{fig:functionalforms_1}
\end{figure}

\subsection{Validation of the versatility of the proposed estimator}

\begin{figure}
	\centering
	\includegraphics[width=0.5\textwidth]{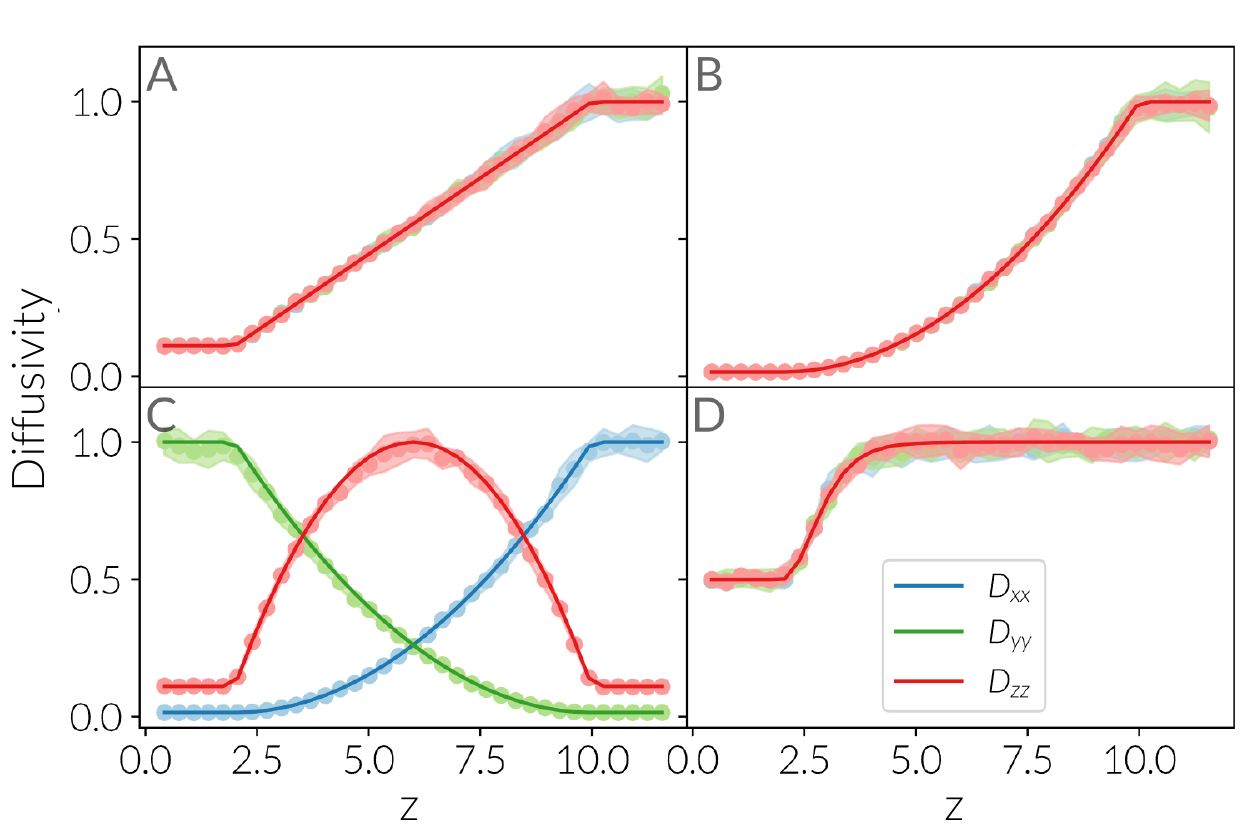}
	\caption{Diffusivity profiles of Fig.~\ref{fig:functionalforms_1} scaled with respect to the maximum value of the corresponding diffusivity component. The error bars have comparable magnitudes suggesting that the relative error is uniform irrespective of the mathematical form and variability of the diffusivity profile.}
	\label{fig:functionalforms_scaled}
\end{figure}

\noindent
We then assess the overall performance of the FCE estimator given by Eq.~(\ref{eq:FCE-estimator-alpha0}) in resolving a number of anisotropic diffusivity profiles with different mathematical forms; namely linear, quadratic, tangent hyperbolic and trigonometric profiles. Fig.~\ref{fig:functionalforms_1} depicts the FCE diffusivity estimates for several non-oscillatory profiles. It can be observed that the FCE estimator is fully capable of accurately capturing the anisotropy and the position dependence of the diffusivity tensor, irrespective of their mathematical form and even when the diffusivity changes by orders of magnitude within the observation domain. Moreover, even though the error bars of the diffusivity estimators might appear to be larger \textcolor{black}{in some components rather than others,} a closer inspection of the relative error reveals that uncertainties are indeed proportional to the local value of the diffusivity. This can, for instance, be vividly observed in Fig.~\ref{fig:functionalforms_scaled}, which depicts $D_{\nu\nu,r}(z)=D_{\nu\nu}(z)/D_{\nu\nu}^{\max}$ vs.~$z$. Here, $\nu\in\{x,y,z\}$ and $D_{\nu\nu}^{\max}$ is the maximum value of $D_{\nu\nu}(z)$ within the observation domain. Indeed, the magnitudes of the error bars are comparable for different scaled profiles, irrespective of the magnitude of $D_{\nu\nu}^{\max}$, which suggests that the relative error of the FCE estimator is fairly uniform and is insensitive to spatial variations of $\textbf{D}(\textbf{r})$.

\begin{figure}
	\centering
	\includegraphics[width=0.5\textwidth]{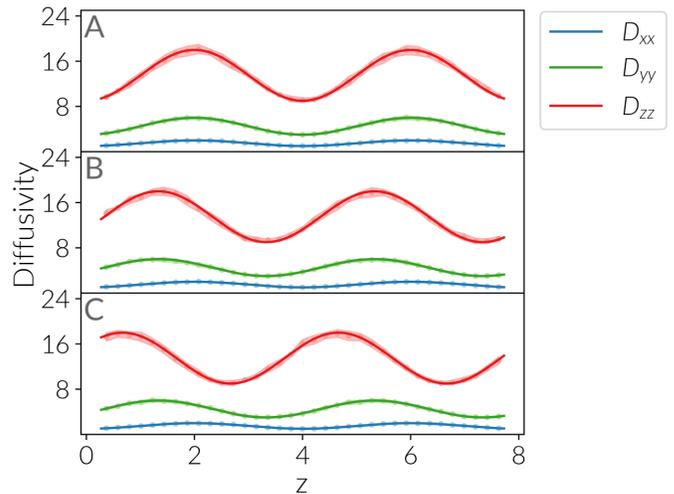}
	\caption{Anisotropic sinusoidal profiles constructed in accordance with Eq.~(\ref{eq:Dz-sin}) with $A_x=1, A_y=3$ and $A_z=9$ and with phases: (A) $\phi_x=\phi_y=\phi_z=0$, (B) $\phi_x=0, \phi_y=\phi_z=\tfrac\pi3$ and (C) $\phi_x=0, \phi_y=\frac{\pi}3, \phi_z=\frac{2\pi}3$.}
	\label{fig:sinprofilephaseshift}
\end{figure}

Despite the diverse mathematical forms of the profiles considered in Figs.~\ref{fig:functionalforms_1} and \ref{fig:functionalforms_scaled}, they are all non-oscillatory. In order to further test the robustness of the proposed FCE estimator, we apply it to trajectories obtained from an oscillatory anisotropic profile of the form:
\begin{eqnarray}
D_{\nu\nu}(z)=A_{\nu}\left[1+\sin^2\left(\frac{\pi z}{4}+\phi_\nu\right)\right] \label{eq:Dz-sin}
\end{eqnarray}
with $A_\nu=1, 3$ and $9$ for $\nu=x, y$ and $z$, respectively. As demonstrated in Fig.~\ref{fig:sinprofilephaseshift}, the FCE estimates are in excellent agreement with the true diffusivity profiles even though the amplitude $A_\nu$ differs by almost an order of magnitude across different components. The agreement is also observed whether all (Fig.~\ref{fig:sinprofilephaseshift}A) and some (Fig.~\ref{fig:sinprofilephaseshift}B) of the diffusivity components are in-phase or if all the components are out-of-phase (Fig.~\ref{fig:sinprofilephaseshift}C).

\begin{figure}
	\centering
	\includegraphics[width=0.5\textwidth]{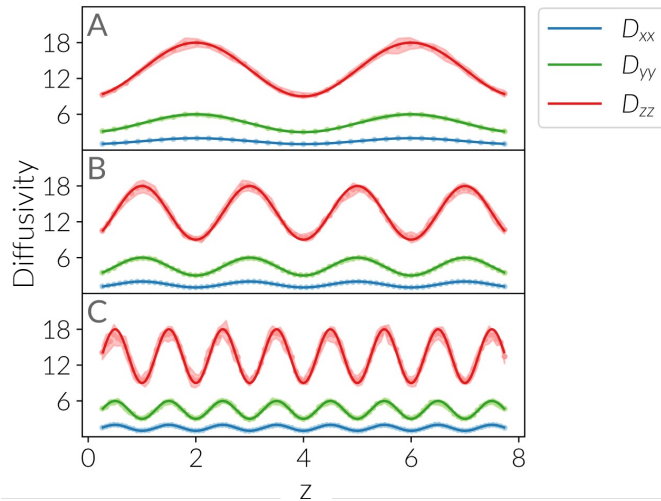}
	\caption{Sinusoidal profiles with increasing wavenumbers. The profiles are of the form $D_{\nu\nu}=A_u\left[1+\sin^2\left(\pi\omega z/8\right)\right]$ with: (A) $\omega=1, \epsilon=0.264$, (B) $\omega=2, \epsilon=0.159$ and (C) $\omega=4, \epsilon=0.053$.}
	\label{fig:sinprofilehighfreq}
\end{figure}

Next, we evaluate the extent by which the FCE estimator can capture rapid oscillations in diffusivity.  It must be noted that low wavelength oscillations in diffusivity make it necessary to use smaller localization parameters $\epsilon$'s. As depicted in Fig.~\ref{fig:sinprofilehighfreq}, there is still a good agreement between the estimates and the true diffusivity profiles. To better illustrate the effect of the width of the localization kernel $\epsilon$, we employ Taylor expansion in terms of $\epsilon$ in Appendix~\ref{appendix2} to demonstrate that:
\begin{eqnarray}
 \widehat{D}_{\textbf{kk}}^{(\epsilon)}-D_{\textbf{kk}}(\textbf{r}_0) &=& \epsilon^2\big[
 \textbf{H}_{D_{\textbf{kk}}}(\textbf{r}_0):\textbf{K}_2\notag\\
 && -2\beta\nabla U(\textbf{r}_0)^T\textbf{K}_2\nabla D_{\textbf{kk}}(\textbf{r}_0)\big]
  + O(\epsilon^3)\notag
 \end{eqnarray}
which directly follows from Eq.~(\ref{eq:localization-error-analysis}) by noting that $\nabla\ln p_0(\textbf{r})=-\beta\nabla U(\textbf{r})$. Here, $\textbf{K}_2$ is the tensorial second moment of $g(\cdot)$ defined in Eq.~(\ref{eq:g-2nd-mom}) and $\textbf{H}_u$ stands for the Hessian of an arbitrary function $u(\textbf{r})$. In the absence of any external field (or free energy profile), the localization error is expected to be strongly correlated with the second derivative of diffusivity. More precisely, $\widehat{D}_{\textbf{kk}}$ is expected to systematically underestimate diffusivity at the maxima of $D_{\textbf{kk}}(z)$ and overestimate it at the minima, due to the negative and positive signs of the second derivative, respectively. As such, the FCE estimator will be less sharp at  extrema. This makes intuitive sense, since the FCE estimator obtains a weighted average of diffusivity within the support of the localization kernel, and such a weighted average will be systematically smaller (or larger) than the maximum (or the minimum) of true diffusivity.

\begin{figure}
	\centering
	\includegraphics[width=0.5\textwidth]{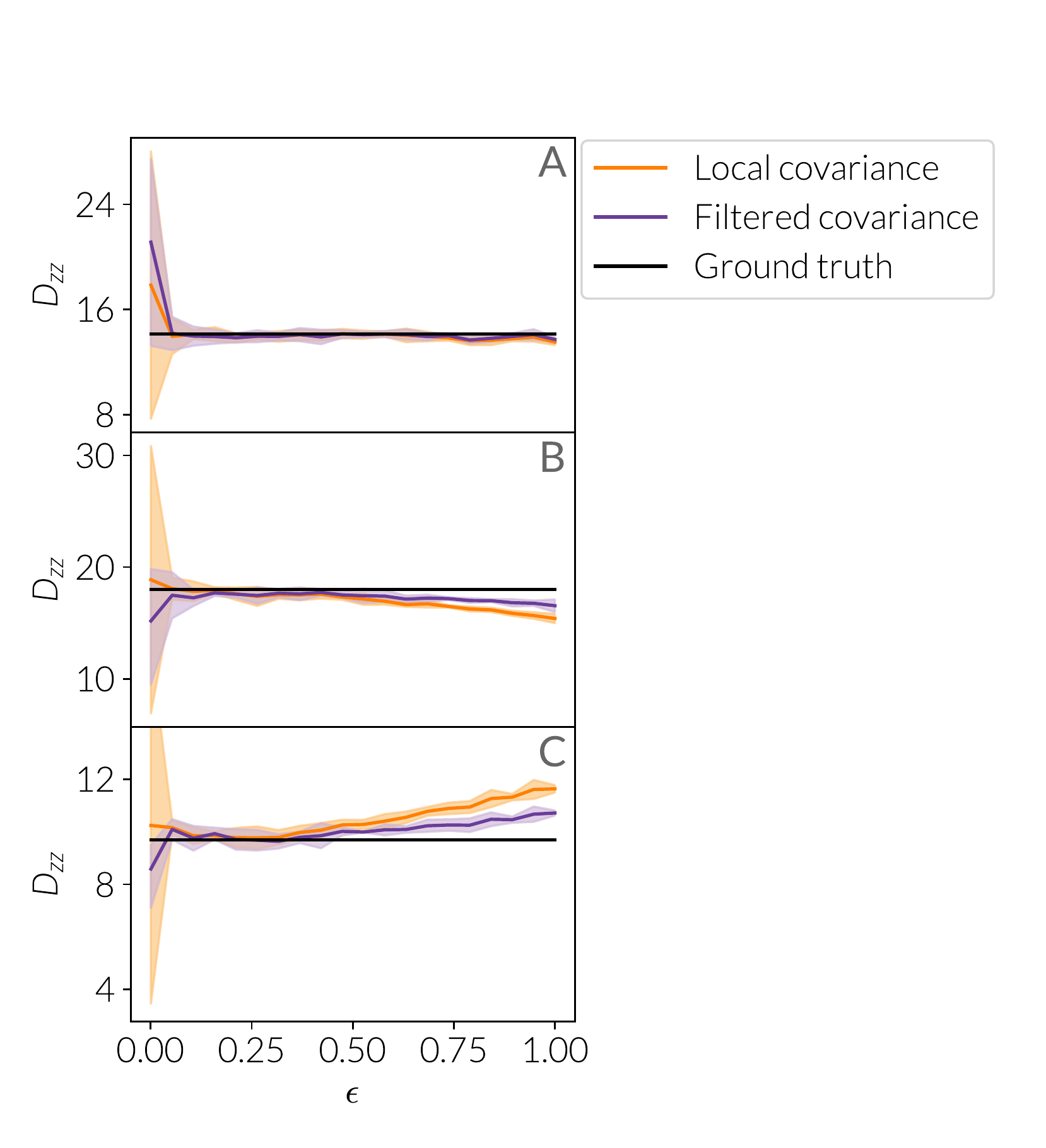}
	\caption{\label{eps_eff_1}The effect of the width of the kernel function $\epsilon$ on the accuracy of the pointwise diffusivity estimate at: (A) an inflection point, (B) a maximum, and (C) a minimum of the diffusivity profile given in Fig.~\ref{fig:sinprofilehighfreq}. In each panel, the true diffusivity, the estimates and the associated uncertainties are depicted in black, solid colored lines and shades, respectively. }	
\end{figure}

 In order to numerically demonstrate this point, we estimate local diffusivity at three fixed points in Fig.~\ref{fig:sinprofilehighfreq}; a maximum, a minimum and an inflection point, but with different values of $\epsilon$. The estimates are depicted in Fig.~\ref{eps_eff_1}. At the maximum and the minimum, increasing $\epsilon$ results in a monotonic deviation from the true diffusivity. This effect is not observed at an inflection point where increasing $\epsilon$ has no discernible impact on the accuracy of the estimate. It must be noted that there is  a tradeoff between mitigating such systematic biases (by decreasing $\epsilon$) and minimizing the sampling uncertainty (by increasing $\epsilon$). Indeed, decreasing $\epsilon$ beyond a certain threshold produces large errors due to insufficient number of particles that are included within the filter as can be seen in Fig.~\ref{eps_eff_1}.
 
We must emphasize that this systematic error is not specific to our FCE estimator and is a fundamental feature of any method that relies on partitioning the observation domain into smaller bins. Indeed, the LCE estimator that is based on defining a generalized MSD function for small bins is more sensitive to $\epsilon$ (i.e.,~the thickness of each bin) than our kernel-based FCE estimator. This is simply because an LCE estimator is a Nadaraya-Watson estimator constructed using the characteristic function of each bin as the kernel, which has a larger second moment $\textbf{K}_2$ than the kernel employed here.

\begin{figure*}
	\centering
	\includegraphics[width=0.9\textwidth]{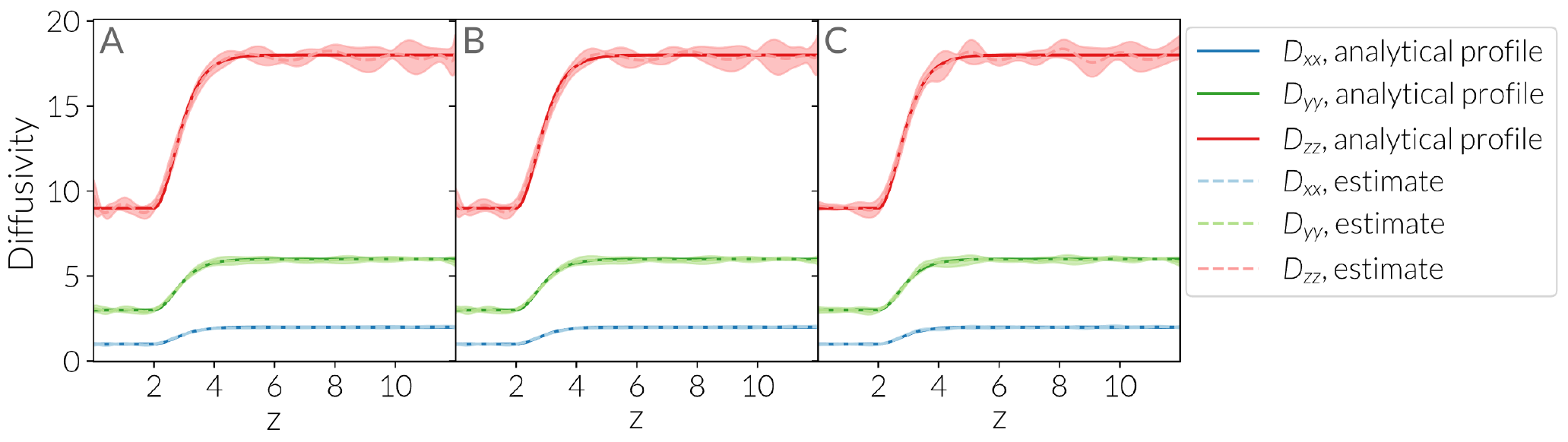}
	\caption{Use of orthogonal basis sets for obtaining the diffusivity profile of Fig.~\ref{fig:functionalforms_1}D. Solid lines correspond to the true diffusivity profiles while the symbols correspond to projection estimates using 15 terms of (A) Legendre and (B) Chebyshev polynomials, and (C) Fourier series.}
	\label{fig:projection}
\end{figure*}

\subsection{Compatibility with orthogonal basis sets}

\noindent
We then numerically test the assertion that the kernel function $g(\cdot)$ does not necessarily have to be localized,~i.e.,~an approximation of the Kronecker delta function. Rather, one can choose a complete basis set of orthogonal functions$\{g_q(\textbf{r})\}_{q=1}^{+\infty}$ and note that $D_{\textbf{kk}}(\textbf{r})$ can be expressed as an infinite sum of  its projections along the elements of the basis set:
$$D_{\textbf{kk}}(\textbf{r})=\sum_{q=1}^{+\infty}c_{q,\textbf{kk}}\,g_q(\textbf{r}).$$ 
with the coefficients of this summation estimated as:
\begin{eqnarray}
c_{n,\textbf{kk}} &=& \left\langle D_{\textbf{kk}}(\textbf{r}), g_n(\textbf{r})
\right\rangle\notag\\
&=& \int D_{\textbf{kk}}(\textbf{r})g_n(\textbf{r})p_0(\textbf{r})\,d^3\textbf{r}\notag\\
&{\color{black}\approx}& {\color{black}\frac{\left\langle\left[g_n(\textbf{X}_{t+h})+g_n(\textbf{X}_{t})\right]
\left[\textbf{k}\cdot\Delta_h\textbf{X}_t\right]^2 \right\rangle_{\textbf{X}_t\sim p_0(\textbf{r})}}{4h}}\notag\\
&&
\label{eq:basis-projection}
\end{eqnarray}
As has been depicted in Fig.~\ref{fig:projection}, a global diffusivity profile can be obtained by using three different basis sets, and the partial sums provide a decent approximation of the true diffusivity profile. This implies that FCE estimators are easier to generalize than their LCE counterparts, in the sense that they can be verifiably formulated to obtain functional estimates of diffusivity (as finite sums of basis functions) rather than a tabulated collection of pointwise diffusivity estimates. In order for this procedure to work, however, it is necessary for the equilibrium probability density $p_0(\textbf{r})$ to be a weight function for the functional inner product employed in (\ref{eq:basis-projection}). In the current study, $p_0(\textbf{r})\equiv\text{const.}$ since particles are chosen from a uniform density profile. As such, conventional basis sets with trivial weight functions, such as Fourier series, can be used without any modification. Applying this approach to systems with non-uniform probability densities (e.g.,~stochastic systems with external fields or molecular dynamics trajectories) will require identifying basis sets with weight functions that emerges from the underlying equilibrium behavior of the system. An alternative approach will be to use the estimator in (\ref{eq:basis-projection}) to compute the projections of $D_{\textbf{kk}}(\textbf{r}) \rho_0(\textbf{r})$ (and not $D_{\textbf{kk}}(\textbf{r})$) onto the elements of a standard basis set, and to divide the arising functional approximation of $D_{\textbf{kk}}(\textbf{r}) \rho_0(\textbf{r})$ by $\rho_0(\textbf{r})$ to obtain $D_{\textbf{kk}}(\textbf{r})$. This procedure might, however, lead to numerical instabilities particularly when $\rho_0(\textbf{r})$ is very small at parts of the observation domain. 


\begin{figure*}
	\centering
	\includegraphics[width=0.75\textwidth]{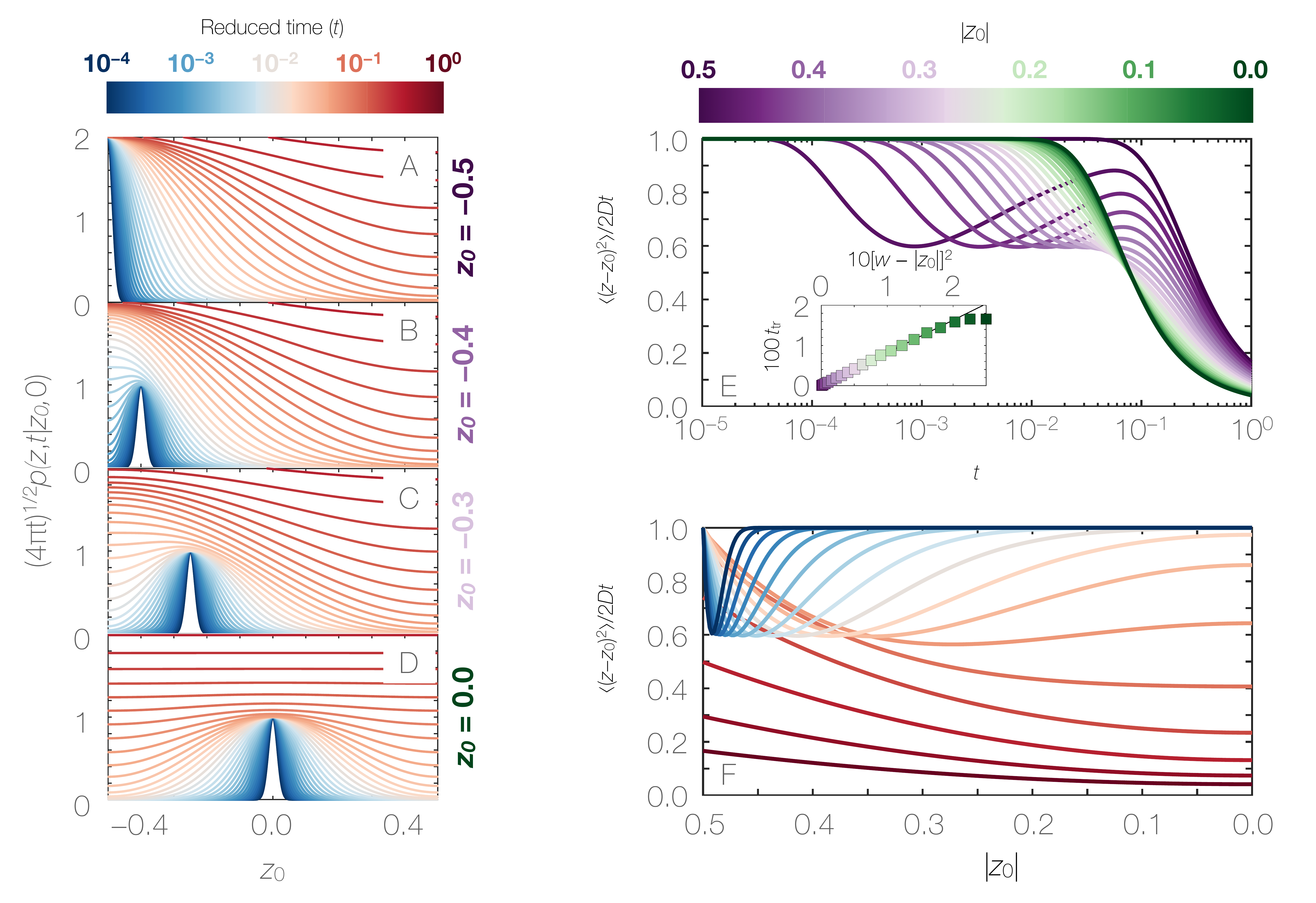}
	\caption{(A-D) The solution of Eq.~(\ref{eq:reflective}) for $w=0.5$ and $D=1$ at (A) $z_0=-0.5$, (B) $z_0=-0.4$, (C) $z_0=-0.3$ and (D) $z_0=0$. Reduced time refers to $tD/4w^2$. (E) Mean-squared displacement vs. $t$ for particles initiated at different $z_0$'s. For $z_0\neq\pm w$,  $t_{\text{tr}}$, the time at which MSD falls below $0.98Dt$ for the first time, exhibits a quadratic scaling with $w-|z_0|$ as can be in the inset. (F) Diffusivity estimate obtained from MSD for different observation widows. The colormap for observation time matches the one used in (A-D).\label{fig:FK-reflective} }
\end{figure*}

\subsection{Hard boundary artifacts}
\label{section:hard-boundary}

\noindent
Intuitively, the temporal discretization at the heart of FCE and LCE estimators rely on the assertion that $p(\textbf{r},t|\textbf{r}_0,0)$ is expected to remain `almost' Gaussian for sufficiently small $t$'s. What constitutes sufficiently small, however, can depend on the nature of confinement and the existence-- or lack thereof-- strong drifts.  As an example, consider one-dimensional diffusion with fixed diffusivity $D$  within the domain $|z|\le w$ that has two reflective boundaries at $z=\pm w$. Such reflective behavior will, for instance, arise in the presence of hard walls that repel an approaching particle with a diverging force.  The solution of the Fokker-Planck equation in such a scenario will no longer be Gaussian and will instead be given by the following Fourier series that can be readily obtained using the separation of variables approach:
\begin{eqnarray}
&&p(z,t|z_0,0)=\frac{1}{2w}\Bigg[1+\sum_{n=1}^{+\infty}e^{-\frac{n^2\pi^2Dt}{w^2}} \cos\frac{n\pi z}{w}\cos\frac{n\pi z_0}{w}\notag\\
&&+\sum_{n=0}^{+\infty}e^{-\frac{(2n+1)^2\pi^2Dt}{4w^2}} \sin\frac{(2n+1)\pi z}{2w}\sin\frac{(2n+1)\pi z_0}{2w}\Bigg]\notag\\
&& \label{eq:reflective}
\end{eqnarray} 
Figs.~\ref{fig:FK-reflective}A-D depict the solution of Eq.~(\ref{eq:reflective}) for $D=1$ and $w=0.5$. As can be seen in Figs.~\ref{fig:FK-reflective}B-D, for  particles that do not start at the boundary (i.e.,~for $|z_0|<w$), $p(z,t|z_0,0)$ always starts as a Gaussian, but eventually  deforms as the probability cloud collides with the closer reflective boundary. The case of starting right the reflective boundary (i.e.,~$z_0=\pm w$) is unique in the sense that $p(z,t|0,0)$ is never Gaussian but rather resembles a folded normal distribution for small $t$'s. Nonetheless, it also gets deformed upon colliding with the second reflective boundary as shown in Fig.~\ref{fig:FK-reflective}A. 

\begin{figure}
	\centering
	\includegraphics[width=.45\textwidth]{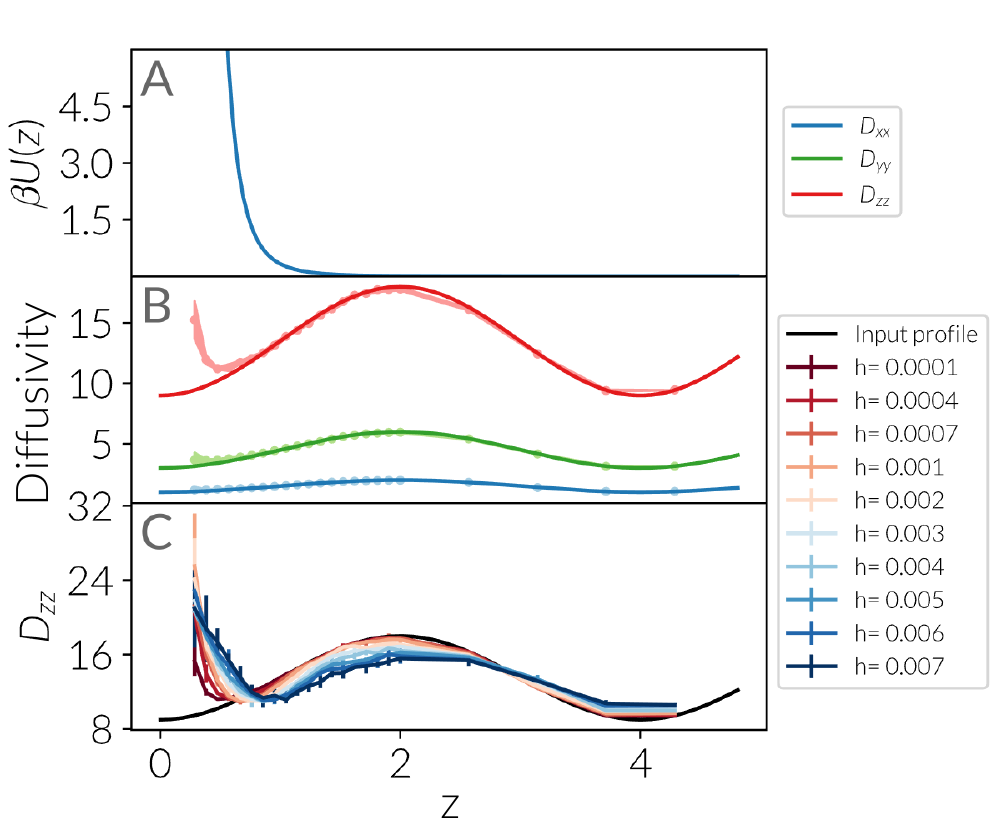}
	\caption{\label{fig:drift}(A) The potential energy landscape given by Eq.~(\ref{eq:drift}) experienced by a particle that moves in accordance with the anisotropic diffusivity profile depicted in solid lines in (B). The estimates obtained from the FCE estimator with $h=10^{-4}$ and the associated uncertainties are depicted in (B) by circles and shades, respectively. In order  to mitigate hard boundary artifacts, a considerably smaller $h$ needs to be employed as shown in (C).  }
	\vspace{-15pt}
\end{figure}

Such deviations adversely impact the performance of covariance-based estimators, first and foremost the mean-squared displacement. For fixed diffusivity and unbounded observation domains, it can be shown that $\langle\left[z(t)-z_0\right]^2\rangle_{z_0} /2Dt = 1$. 
As demonstrated in Fig.~\ref{fig:FK-reflective}E, however, $\langle\left[z(t)-z_0\right]^2\rangle_{z_0}/2Dt$ starts deviating from unity at a timescale that is comparable in magnitude to
\begin{eqnarray}
t_c&\sim&\frac{\left(w-|z_0|\right)^2}{D},
\label{eq:t-tr}
\end{eqnarray}
which corresponds to the characteristic time that it takes for a tracker to diffuse by a distance $w-|z_0|$.  Indeed, there is a quadratic scaling between $t_{\text{tr}}$, the time at which $\langle\left[z(t)-z_0\right]^2\rangle_{z_0}/2Dt$ drops below $0.98$ for the first time, and $w-|z_0|$ as can be seen in the inset of Fig.~\ref{fig:FK-reflective}E. Once again, the case of $z_0=\pm w$ is special in the sense that $t_\text{tr}$ is considerably longer since deviations only arise when the folded normal cloud collides with the other wall. Consistent with this observation, $t_{\text{tr}}|_{z_0=w}$ (not shown in Fig.~\ref{fig:FK-reflective}E) is almost four times larger than $t_{\text{tr}}|_{z_0=0}$ as the folded normal cloud has to travel twice as long to reach the other boundary.

When it comes to a fixed observation window $t$, deviations of $\langle\left[z(t)-z_0\right]^2\rangle_{z_0}/2Dt$ from unity are not spatially uniform as shown in Fig.~\ref{fig:FK-reflective}F. For short times, such deviations are only confined to the immediate vicinity of the reflective boundaries, but they then propagate to the entire domain as $t$ increases. One can expect similar artifacts for other covariance-based estimators (including the FCE introduced in this work) when it comes to estimating the component of the diffusivity tensor normal to a hard boundary. Such artifacts are, however, likely to be mitigated by choosing a sufficiently small observation window (as suggested by Fig.~\ref{fig:FK-reflective}F).

In order to assess the significance of such artifacts for more realistic scenarios,  we generated SDE trajectories according to an oscillatory diffusivity profile (solid lines of Fig.~\ref{fig:drift}B) but under the influence of the potential energy landscape  given by
\begin{eqnarray}
\beta U(z) &=& \frac{1}{3}\left(\frac{1}{z^5} + \frac{1}{\left(L-z\right)^5}\right),~~~0<z<L
\label{eq:drift}
\end{eqnarray}
and shown in Fig.~\ref{fig:drift}A. This potential energy profiles mimics the existence of two stiff boundaries at $z=0, L$. As demonstrated in Fig.~\ref{fig:drift}B, using an observation window of $h=10^{-4}$ leads to considerable systematic errors in $D_{zz}$, the component of the diffusivity tensor perpendicular to the stiff boundary. No such artifact is observed for the tangential components of diffusivity, namely $D_{xx}$ and $D_{yy}$. The magnitude of such systematic errors can, however, be effectively mitigated upon choosing a smaller $h$ as shown in Fig.~\ref{fig:drift}C. Indeed choosing an observation window of $h=10^{-5}$ resolves these artifacts for $z$'s at which sufficient statistics is available,~i.e.,~within parts of the domain at which  $\beta U(z)\lesssim O(1)$. 

\begin{figure}
{\color{black}
	\centering
	\includegraphics[width=.45\textwidth]{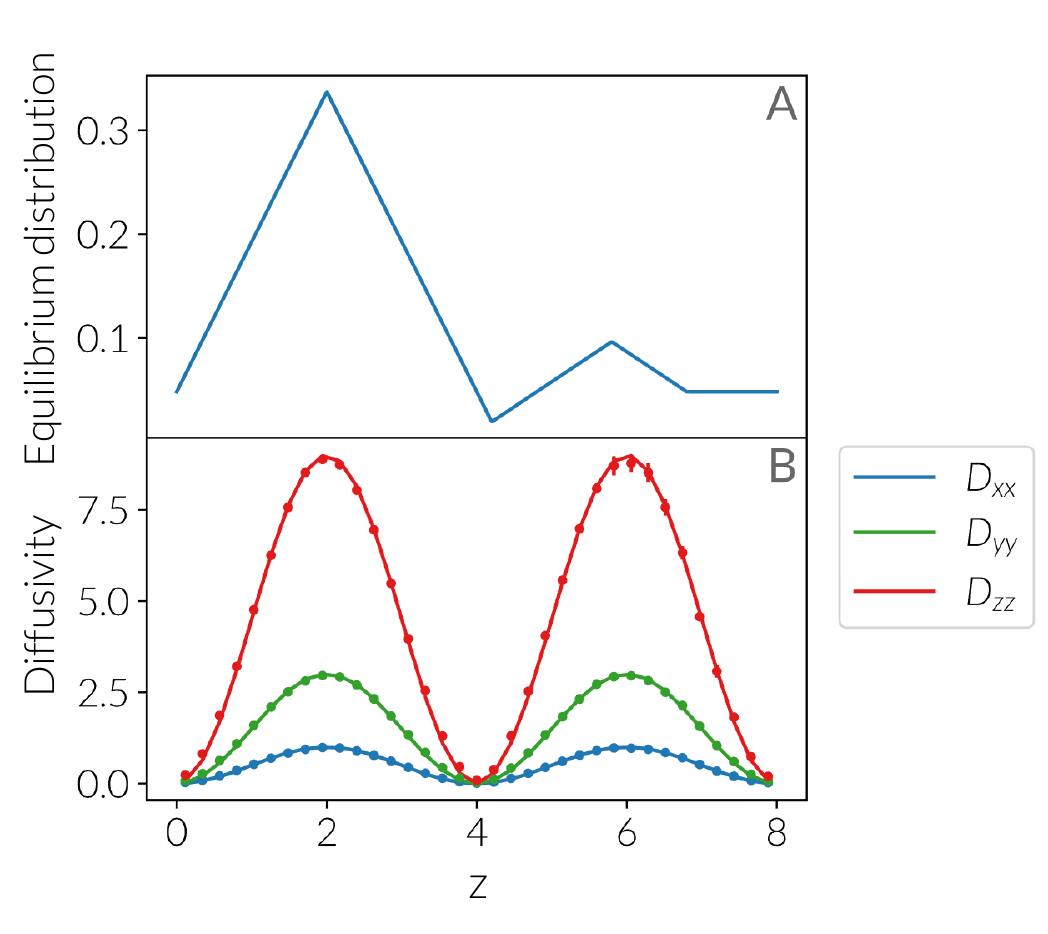}
	\caption{The performance of the FCE estimator when $p_0(\textbf{r})$ is not uniform but lacks diverging drifts, such as the one depicted in (A). The diffusivity estimates obtained using (\ref{eq:FCE-estimator-alpha0}) perfectly match with the actual diffusivity.\label{fig:nonuniform}}
}
\end{figure}

{\color{black}
It must be emphasized that these artifacts are only limited to situations in which the drift term diverges at a particular boundary. Otherwise, if changes in number density (or concentration) are mild,  the estimators given by (\ref{eq:FCE-estimator}) and (\ref{eq:FCE-estimator-alpha0}) will still be capable of capturing spatial variations in diffusivity only with a slight decrease in the observation window, $h$.  In order to test this assertion numerically, we generate SDE trajectories in which particles adopt a non-uniform number density profile given by Fig.~\ref{fig:nonuniform}A, corresponding to a mild barrier of $3kT$ between $z\approx2$ and $z\approx4$. As can be seen in Fig.~\ref{fig:nonuniform}B, the estimator given by (\ref{eq:FCE-estimator-alpha0}) accurately captures oscillations in diffusivity without a need to decrease $h$. This simple test reveals that the FCE estimators proposed here are general and can be readily used in most realistic circumstances where particle density or concentration is not spatially uniform but undergoes mild spatial variations.
}

\section{Conclusions}
\label{section:conclusions}

\noindent
In this work, we present {\color{black}two closely related} filtered covariance estimators for computing position-dependent anisotropic diffusivity profiles from single-particle trajectories. We assess the performance of {\color{black} these estimators by applying them} to trajectories generated according to (\ref{Overdamped Langevin}) with known diffusivity profiles. Overall, the {\color{black}estimators perform} reasonably well in accurately capturing the spatial variations and the anisotropy of the diffusivity tensor even when it changes by several orders of magnitude within the observation domain or when its different components differ by several orders of magnitude at a particular point. We also demonstrate that the estimators are very robust to the particular choice of the kernel function and the localization parameter. We, however, observe systematic errors in estimating normal components of diffusivity at the vicinity of stiff boundaries. Luckily, such errors can be mitigated by choosing a shorter observation window. {\color{black}Finally, we find the $\alpha\rightarrow0$ estimator to be more robust to numerical instabilities caused by poorly conditioned stencils.}

It must be noted that an approach similar to ours can be employed to construct more potent LCE estimators,~i.e.,~by using a localization kernel to define a more localized $p_1(\cdot)$ in (\ref{eq:int-D-confined}). Moreover, the covariance at the right hand side of Eqs.~(\ref{eq:int-D-confined}) and (\ref{eq:LCE-estimator}) can be modified by projecting displacement along a stencil of unit vectors. We, however, contend that the FCE estimator introduced here will still be  superior to such LCE generalizations since it spatially resolves the position of the particle both at the beginning and at the end of the observation window, something that is not feasible even with its LCE-based counterparts.

The FCE {\color{black}estimators} introduced here {\color{black}are} prone to two types of errors. The first is the localization error, or the error due to the finite thickness of the localization kernel. We demonstrate that the localization error might systematically bias the  estimates of diffusivity at the maxima and minima, or when there is a strong external force. While using a narrower kernel will mitigate this bias, decreasing the width of the kernel beyond a certain level can adversely impact the variance of the estimator. As such, the task of identifying an optimal $\epsilon$ might require solving an optimization problem.

The second source of uncertainty is the discretization error, which arises due to the finite duration of the observation window,~i.e.,~the timeframe over which the displacement of the particle is processed by the filter function. As discussed in Section~\ref{section:hard-boundary}, such discretization errors can be particularly significant in the presence of stiff boundaries with divergent repelling forces. {\color{black}They are, however, generally minimal if changes in particle density or concentration are mild.}  While discretization errors will generally decrease upon letting $h\rightarrow0^+$, there might be a lower practical limit for $h$,~e.g.,~due to the limited temporal resolution of experiments, or the fundamental inability of Eq.~(\ref{Overdamped Langevin}) to describe the behavior of single-particle displacements over short timescales. We, however, describe a framework in Appendix~\ref{appendix3} to estimate the extent of the discretization error by solving the adjoint Smoluchowski equation analytically or numerically.  As to how to detect and solve such artifacts is beyond the focus of this paper and will be discussed in our follow-up paper.

One of the issues that was not addressed in this work is the corrections due to measurement errors. While this is not an issue of concern in trajectories attained from simulations, it is something that can impact the applicability of the estimator to experimental data. Another issue that was not probed here is the possibility of many-body effects (e.g.,~correlated motion of many particles), which might require introducing corrections to the proposed FCE estimator. Both these issues will be topics of future explorations. Despite these limitations, however, the estimator proposed in this work is a useful contribution to the problem of estimating diffusivity profiles from single-particle trajectories given its simplicity and flexibility (particularly its compatibility to be used with a dictionary of orthogonal functions). Moreover, while {\color{black}these estimators were} developed to compute diffusivity from actual single-particle trajectories, {\color{black}they} can in principle be applied to quantify diffusivity in collective variable spaces, which can also evolve according to overdamped Langevin dynamics.

\appendix

\section{Derivation of Eq.~(\ref{AverGammaC2})}\label{appendix1}

\noindent
Consider the functions $\gamma_{\pm}(\textbf{r}):=f_{\textbf{k}}(\textbf{r})\left[G(\textbf{r})\pm\frac{1}{4}\right] = \gamma_{\textbf{k}}(\textbf{r}) \pm \tfrac14f_{\textbf{k}}(\textbf{r})$ and note that that:
\begin{eqnarray}
\nabla\gamma_{\pm}(\textbf{r}) &=& \nabla f_{\textbf{k}}\left[G(\textbf{r})\pm\frac14\right] + f_{\textbf{k}}(\textbf{r})\nabla G\notag\\
&=& e^{-i\alpha\textbf{k}\cdot\textbf{r}}\left\{
-i\alpha\textbf{k}\left[G(\textbf{r})\pm\frac14\right] +\nabla G
\right\}
\end{eqnarray}
which yields:
\begin{eqnarray}
&& \nabla \gamma^{\dagger}_{\pm}\textbf{D}(\mathbf{r})\nabla\gamma_{\pm}  = \left\{
i\alpha\textbf{k}\left[G(\textbf{r})\pm\frac14\right] +\nabla G
\right\}^T\textbf{D}(\textbf{r})\notag\\
&& \times \left\{
-i\alpha\textbf{k}\left[G(\textbf{r})\pm\frac14\right] +\nabla G
\right\}\notag\\
&& = \alpha^2 \textbf{k}^T\textbf{D}\textbf{k}\left[
G(\textbf{r})\pm\frac14
\right]^2 + \nabla G^T\textbf{D}\nabla G \notag\\
&& +i\alpha\left[
G(\textbf{r})\pm\frac14
\right]\left[
\textbf{k}^T\textbf{D}\nabla G-\nabla G^T\textbf{D}\textbf{k}
\right]\notag
\end{eqnarray}
 However, the last term vanishes because $\textbf{D}$ is symmetric according to Onsager's reciprocity principle. 
The left hand side of Eq.~\eqref{AverGammaC}  will therefore be given by:
\begin{eqnarray}
&& \int \nabla \gamma^{\dagger}_{\pm}\textbf{D}(\mathbf{r})\nabla\gamma_{\pm} p_0d^3\mathbf{r} = \int \nabla G^T\textbf{D}\nabla Gp_0(\textbf{r}) d^3\textbf{r}\notag\\
&& +\alpha^2\int\textbf{k}^T\textbf{D}\textbf{k}\left[
G(\textbf{r})\pm\frac14
\right]^2p_0(\textbf{r})d^3\textbf{r}
\end{eqnarray}
The left hand side of Eq.~(\ref{AverGammaC2}) can thus be obtained through the following subtraction:
\begin{eqnarray}
&&\int \nabla \gamma^{\dagger}_{+}\textbf{D}(\mathbf{r})\nabla\gamma_{+} p_0d^3\mathbf{r}-\int \nabla \gamma^{\dagger}_{-}\textbf{D}(\mathbf{r})\nabla\gamma_{-} p_0d^3\mathbf{r}\notag\\
&& = \alpha^2\int\textbf{k}^T\textbf{D}\textbf{k}\left[
G(\textbf{r})+\frac14
\right]^2p_0(\textbf{r})d^3\textbf{r}\notag\\
&& - \alpha^2\int\textbf{k}^T\textbf{D}\textbf{k}\left[
G(\textbf{r})-\frac14
\right]^2p_0(\textbf{r})d^3\textbf{r}\notag\\
&& = 
\alpha^2\int \textbf{k}^T\textbf{D}\textbf{k}G(\textbf{r})p_0(\textbf{r})d^3\textbf{r}
\notag
\end{eqnarray}
When it comes to the right hand side of Eq.~\eqref{AverGammaC}, it can be expressed as:
\begin{eqnarray}
&& \left\langle 
\left|\gamma_{\pm}(\textbf{X}_{t+h}) - \gamma_{\pm}(\textbf{X}_{t})
\right|^2
\right\rangle \notag\\
&& = \left\langle \left|\gamma_{\textbf{k}}(\textbf{X}_{t+h}) - \gamma_{\textbf{k}}(\textbf{X}_{t})
\pm\frac{
f_{\textbf{k}}(\textbf{X}_{t+h}) - f_{\textbf{k}}(\textbf{X}_{t})}{4}
\right|^2
\right\rangle \notag\\
&& = \left\langle
 \left|\gamma_{\textbf{k}}(\textbf{X}_{t+h}) - \gamma_{\textbf{k}}(\textbf{X}_{t})\right|^2
\right\rangle+\frac{\left\langle
 \left|
 f_{\textbf{k}}(\textbf{X}_{t+h}) - f_{\textbf{k}}(\textbf{X}_{t})
 \right|^2
 \right\rangle}{16}\notag\\
 && \pm\frac12\Re \left\langle
 \left[
 \gamma_{\textbf{k}}(\textbf{X}_{t+h}) - \gamma_{\textbf{k}}(\textbf{X}_{t})
 \right]^*
 \left[
 f_{\textbf{k}}(\textbf{X}_{t+h}) - f_{\textbf{k}}(\textbf{X}_{t})
 \right]
 \right\rangle\notag
\end{eqnarray}
Subtracting the two expectations completed the proof for  Eq.~(\ref{AverGammaC2}).

{\color{black}
\section{Derivation of Eq.~(\ref{eq:FCE-estimator-alpha0}), the FCE estimator at the limit $\alpha\rightarrow0$}
\label{appendix:alpha0}

\noindent 
It can be readily demonstrated that:
\begin{eqnarray}
 \Delta_h\gamma_{\textbf{k}}^*(\textbf{X}_{i,t})\Delta_hf_{\textbf{k}}^*(\textbf{X}_{i,t}) &=& \left[1-e^{i\alpha\textbf{k}\cdot\Delta_h\textbf{X}_{i,t}}\right]G(\textbf{X}_{i,t+h}) \notag\\ 
 && +  \left[1-e^{-i\alpha\textbf{k}\cdot\Delta_h\textbf{X}_{i,t}}\right]G(\textbf{X}_{i,t}) 
 \notag
\end{eqnarray}
As such, the real part of $\Delta_h\gamma_{\textbf{k}}^*(\textbf{X}_{i,t})\Delta_hf_{\textbf{k}}^*(\textbf{X}_{i,t})$ is given by:
\begin{eqnarray}
&&\Re\left[\Delta_h\gamma_{\textbf{k}}^*(\textbf{X}_{i,t})\Delta_hf_{\textbf{k}}^*(\textbf{X}_{i,t})\right] =\notag\\
&& \left[
1-\cos\left(
\alpha\textbf{k}\cdot\Delta_h\textbf{X}_{i,t}
\right)
\right]\left[G(\textbf{X}_{i,t+h})+G(\textbf{X}_{i,t})\right]
\notag
\end{eqnarray}
The proof readily follows by noting that $\lim_{x\rightarrow0} [1-\cos ax]/x^2=a^2/2$.
}

\section{Asymptotic properties of the estimator}

\subsection{The effect of the localization parameter $\epsilon$}\label{appendix2}

\noindent The expression on the right hand side of Eq.~\eqref{Localized_gammaC} can be cast as:
\[\hat{D}_{\textbf{kk}}=\int \textbf{k}^{T}\textbf{D}(\mathbf{r})\textbf{k}\, \omega_\epsilon(\textbf{r},\textbf{r}_0)d^3\mathbf{r}\]
where $\omega_\epsilon$ is a localization function weighted by the stationary measure $p_0$:
\[\omega_\epsilon(\textbf{r},\textbf{r}_0)=\frac{p_0(\textbf{r})g_\epsilon(\mathbf{r}-\textbf{r}_0)}{\int p_0(\textbf{s})g_\epsilon(\mathbf{s}-\textbf{r}_0)\,d^3\textbf{s}}\]
Here, $g_\epsilon(\textbf{r})=\epsilon^{-d}g(\epsilon^{-1}\textbf{r})$. 
Suppose that $g_\epsilon(\cdot)$ satisfies the property that
\begin{eqnarray}
 \int (\mathbf{r}-\textbf{r}_0)g_\epsilon(\textbf{r}-\textbf{r}_0)\,d^3\textbf{r}&=&\pmb 0,
  \label{eq:g-1st-mom}
 \end{eqnarray}
 and define:
 \begin{eqnarray}
\mathbf{K}_2&:=&\frac{1}{2}\int \textbf{y}\textbf{y}^Tg(\textbf{y})\, d^3\textbf{y}.
\label{eq:g-2nd-mom}
\end{eqnarray}
Letting $\textbf{r} = \textbf{r}_0+\epsilon\textbf{y}$ and conducting a Taylor expansion around $\textbf{r}_0$ yields:
\begin{eqnarray}
D_{\textbf{kk}}(\textbf{r}) &=& D_{\textbf{kk}}(\textbf{r}_0) + \epsilon\textbf{y}^T\nabla D_{\textbf{kk}}(\textbf{r}_0)\notag\\
&& + \frac{\epsilon^2}{2!}\textbf{H}_{D_{\textbf{kk}}}(\textbf{r}_0):\textbf{y}\textbf{y}^T + O(\epsilon^3)\\
p_0(\textbf{r}) &=& p_0(\textbf{r}_0) + \epsilon\textbf{y}^T\nabla p_0(\textbf{r}_0) \notag\\
&& + \frac{\epsilon^2}{2!}\textbf{H}_{p_0}(\textbf{r}_0):\textbf{y}\textbf{y}^T + O(\epsilon^3)\label{eq:p0-Taylor}
\end{eqnarray}
where $\textbf{H}_{D_{\textbf{kk}}}(\textbf{r}_0)$ and $\textbf{H}_{p_0}(\textbf{r}_0)$ are the Hessians of $D_{\textbf{kk}}(\textbf{r})$ and $p_0(\textbf{r})$ estimated at $\textbf{r}_0$, respectively, and $\textbf{A}:\textbf{B}$ corresponds to full  contraction between tensors $\textbf{A}$ and $\textbf{B}$. For the denominator, one can use (\ref{eq:g-1st-mom}), (\ref{eq:g-2nd-mom}) and (\ref{eq:p0-Taylor}) to demonstrate that:
\begin{eqnarray}
\int p_0(\textbf{s})g_\epsilon(\textbf{s}-\textbf{r}_0)d^3\textbf{s} &=&  p_0(\textbf{r}_0)+ \epsilon^2 {\textbf{H}_{p_0}(\textbf{r}_0):\textbf{K}_2}+ O(\epsilon^3) 
\notag\\ && 
\end{eqnarray}
Similarly, $\widehat{D}_{\textbf{kk}}$ can  be expressed as:
\begin{eqnarray}
\widehat{D}_{\textbf{kk}} &=& \int d^3\textbf{y}\,g(\textbf{y})\Bigg\{ D_{\textbf{kk}}(\textbf{r}_0)p_0(\textbf{r}_0)+\notag\\
&& \epsilon\textbf{y}^T\left[
D_{\textbf{kk}}(\textbf{r}_0)\nabla p_0(\textbf{r}_0) + p_0(\textbf{r}_0)\nabla D_{\textbf{kk}}(\textbf{r}_0)
\right]+\notag\\
&& \frac{\epsilon^2}{2}\bigg[
p_0(\textbf{r}_0)\textbf{H}_{D_\textbf{kk}}(\textbf{r}_0) + D_{\textbf{kk}}(\textbf{r}_0)\textbf{H}_{p_0}(\textbf{r}_0) + \notag\\
&& 2\nabla D_{\textbf{kk}}(\textbf{r}_0)\nabla p_0^T(\textbf{r}_0)
\bigg]:\textbf{y}\textbf{y}^T + O(\epsilon^3)\Bigg\}\times
\notag\\
&& \left\{
\frac{1}{p_0(\textbf{r}_0)} - \frac{\epsilon^2\textbf{H}_{p_0}(\textbf{r}_0):\textbf{K}_2}{p_0^2(\textbf{r}_0)} + O(\epsilon^4)
\right\}\notag\\
&=& D_{\textbf{kk}}(\textbf{r}_0) - \frac{\epsilon^2D_{\textbf{kk}}(\textbf{r}_0)\textbf{H}_{p_0}(\textbf{r}_0):\textbf{K}_2}{p_0(\textbf{r}_0)} \notag\\
&& + \frac{\epsilon^2}{2}\Big[\textbf{H}_{D_\textbf{kk}}(\textbf{r}_0)
 + \frac{D_{\textbf{kk}}(\textbf{r}_0)\textbf{H}_{p_0}(\textbf{r}_0)}{p_0(\textbf{r}_0)}\notag\\
 &&
 + \frac{2\nabla D_{\textbf{kk}}(\textbf{r}_0)\nabla p_0^T(\textbf{r}_0)}{p_0(\textbf{r}_0)}
 \Big]:\int \textbf{y}\textbf{y}^Tg(\textbf{y})\,d^3\textbf{y} \notag\\
 &=& D_{\textbf{kk}}(\textbf{r}_0) + \frac{\epsilon^2}{p_0(\textbf{r}_0)} \bigg[
 p_0(\textbf{r}_0)\textbf{H}_{D_{\textbf{kk}}}(\textbf{r}_0)\notag\\
 && +2\nabla D_{\textbf{kk}}(\textbf{r}_0)\nabla p_0^T(\textbf{r}_0)
 \bigg]:\textbf{K}_2 + O(\epsilon^3)
 \label{eq:localization-error-analysis}
\end{eqnarray}
This proves that the localization {\color{black}errors of the FCEs described in Algorithms~\ref{alg:fce} and \ref{alg:fce-alpha0}  scale} with $\epsilon^2$. Similar to the theory of differential equations, one can obtain more accurate estimators by eliminating the desired powers of $\epsilon$ via constructing linear combinations of estimators (each using a slightly different $\epsilon$). For instance, the leading localization error term in the following estimator will be $O(\epsilon^3)$: 
\begin{eqnarray}{\label{2nd order correction}}
\widehat{D}^{\text{3rd-order}}_{\textbf{kk}} &=& 
	2\widehat{D}^{\epsilon}_{\textbf{kk}}-D^{\sqrt{2}\epsilon}_{\textbf{kk}}
\end{eqnarray}
It is also possible to compute $\widehat{D}^{\epsilon}_{\textbf{kk}}(\textbf{r}_0)$ for different $\epsilon$ values and then perform a quadratic regression to extrapolate to the limit of $\epsilon\rightarrow0$. It must, however, be noted that reducing $\epsilon$-- beyond a certain level-- can adversely impact the performance of the estimator due to lack of proper statistics.

\subsection{The effect of time discretization}\label{appendix3}

\noindent
The second question that we wish to address is how temporal discretization (i.e.,~using a finite $h$) impacts the error in the estimator. In order to analyze such errors, one can observe that the right hand side of the FCE estimator {\color{black}given by Eqs.}~\eqref{AverGammaC2} and ~\eqref{AverGammaC2-alpha0} can be expressed in terms of $p_\tau(\textbf{s}|\textbf{r})\equiv p(\textbf{X}_\tau=\textbf{s}|\textbf{X}_0=\textbf{r}){\color{black}=e^{\tau\mathcal{L}_{\mathbf{s}}}\delta(\mathbf{s}-\mathbf{r})}$, which is also known as the self-part of the van Hove correlation function.\cite{vanHovePhysRev1954} {\color{black}Here, $\mathcal{L}_{\textbf{s}}\equiv\nabla_{\textbf{s}}\cdot\left[\textbf{D}(\textbf{s})\cdot\left(\nabla_{\textbf{s}}+\beta\nabla_{\textbf{s}} U(\textbf{s})\right)\right]$  is the differential operator in the right hand side of (\ref{Smoluch}). Now suppose that $F:\mathbb{R}^d\rightarrow\mathbb{R}^d\rightarrow\mathbb{C}$ is an integrable smooth function such that $F(\textbf{x},\textbf{x})=0$. It can thus be demonstrated that:
\begin{eqnarray}
	&&\lim\limits_{\tau\rightarrow 0^+}\frac{1}{\tau}\left\langle F(\mathbf{X}_{t+\tau},\mathbf{X}_t)\right\rangle_{\mathbf{X}_t\sim 
	p_0(\mathbf{r})}=\notag\\
	&&\lim\limits_{\tau\rightarrow 0^+}\frac{1}{\tau}\int\int F(\mathbf{s},\mathbf{r}) p_\tau(\mathbf{s}|\mathbf{r})p_0(\mathbf{r})\,d\mathbf{s}\,d\mathbf{r}=\notag\\
	&&\lim\limits_{\tau\rightarrow 0^+}\frac{1}{\tau}\int\int F(\mathbf{s},\mathbf{r}) e^{\tau\mathcal{L}_\mathbf{s}}\left[\delta(\mathbf{s}-\mathbf{r})\right]p_0(\mathbf{r})\,d\mathbf{s}\,d\mathbf{r}=\notag\\
	&&\lim\limits_{\tau\rightarrow 0^+}\int\int F(\mathbf{s},\mathbf{r}) \left\{\left[
	\frac{e^{\tau\mathcal{L}_\mathbf{s}}-1}{\tau}
	\right] \delta(\mathbf{s}-\mathbf{r})\right\}
	p_0(\mathbf{r})\,d\mathbf{s}\,d\mathbf{r}\notag\\
	&&=\int\int F(\mathbf{s},\mathbf{r}) \mathcal{L}_\mathbf{s}\left[\delta(\mathbf{s}-\mathbf{r})\right]p_0(\mathbf{r})\,d\mathbf{s}\,d\mathbf{r}\notag\\
	&&=\int\int \mathcal{L}_\mathbf{s}^\dagger\left[F(\mathbf{s},\mathbf{r})\right] \delta(\mathbf{s}-\mathbf{r})p_0(\mathbf{r})\,d\mathbf{s}\,d\mathbf{r}\label{eq:appendix:Fxy-exp}
\end{eqnarray}
Here, $\mathcal{L}^{\dagger}[u(\textbf{r},\textbf{s})] \equiv-\beta(\nabla_{\textbf{s}}U)^T \textbf{D} \nabla_{\textbf{s}}u+ (\nabla_{\textbf{s}}\cdot\textbf{D})\cdot\nabla_{\textbf{s}}u+\textbf{D}:\textbf{H}_{\textbf{s},u}$  is the adjoint operator of $\mathcal{L}_{\textbf{s}}$.  The task of evaluating Eq.~(\ref{eq:appendix:Fxy-exp}) can therefore be reduced for identifying the gradient and the Hessian of $F$. For the $\alpha\neq0$ estimator of Eq.~\eqref{AverGammaC2}, $F(\textbf{s},\textbf{r})$ is given by:
\begin{eqnarray}
F(\textbf{s},\textbf{r}) &=& \frac{\left[f_\mathbf{k}(\mathbf{s})g(\mathbf{s}-\mathbf{r}_0)-f_\mathbf{k}(\mathbf{r})g(\mathbf{r}-\mathbf{r}_0)\right]\left[f_\mathbf{k}(\mathbf{s})-f_\mathbf{k}(\mathbf{r})\right]}{2\alpha^2}\notag\\
&&
\end{eqnarray}
It can be noted that:
}
\begin{eqnarray}
&& \mathcal{L}^{\dagger}_{\textbf{s}}\bigg\{\big[f_\textbf{k}(\textbf{s})g(\textbf{s}-\textbf{r}_0) -f_\textbf{k}(\textbf{r})g(\textbf{r}-\textbf{r}_0)\big]^*\left[f_{\textbf{k}}(\textbf{s})-f_{\textbf{k}}(\textbf{r}) \right]\bigg\} \notag\\
&& =  \mathcal{L}^{\dagger}_{\textbf{s}}\left[g(\textbf{s}-\textbf{r}_0)\right] - e^{-i\alpha\textbf{k}\cdot\textbf{r}}\mathcal{L}^{\dagger}_{\textbf{s}}\left[e^{i\alpha\textbf{k}\cdot\textbf{s}}g(\textbf{s}-\textbf{r}_0)\right] \notag\\
&& - e^{i\alpha\textbf{k}\cdot\textbf{r}}g(\textbf{r}-\textbf{r}_0)\mathcal{L}^{\dagger}_{\textbf{s}}\left[e^{-i\alpha\textbf{k}\cdot\textbf{s}}\right]
\label{eq:L-dagger-expand}
\end{eqnarray}
with the application of $\mathcal{L}^{\dagger}_{\textbf{s}}$ to each term given by:
\begin{eqnarray}
\mathcal{L}^{\dagger}_{\textbf{s}}\left[g(\textbf{s}-\textbf{r}_0)\right] &=& (\nabla_{\textbf{s}}\cdot\textbf{D})\cdot\nabla_{\textbf{s}}g + \textbf{D}:\textbf{H}_{\textbf{s},g}\notag\\
&& -\beta(\nabla_{\textbf{s}} U)^T\textbf{D}\nabla_{\textbf{s}}g
\label{eq:L-dagger-g}\\
\mathcal{L}^{\dagger}_{\textbf{s}}\left[e^{i\alpha\textbf{k}\cdot\textbf{s}}g(\textbf{s}-\textbf{r}_0)\right] &=& e^{i\alpha\textbf{k}\cdot\textbf{s}}\big\{(\nabla_{\textbf{s}}\cdot\textbf{D})\cdot\left[i\alpha\textbf{k}g+\nabla_{\textbf{s}}g\right]\notag\\
&& +\textbf{D}:\big[
-\alpha^2\textbf{k}\textbf{k}^Tg + i\alpha\textbf{k}(\nabla_{\textbf{s}}g)^T\notag\\
&& +i\alpha(\nabla_{\textbf{s}}g)\textbf{k}^T+\textbf{H}_{\textbf{s},g}  
\big]\notag\\
&& -\beta(\nabla_{\textbf{s}}U)^T\textbf{D}\left[
i\alpha\textbf{k}g+\nabla_{\textbf{s}}g
\right]\big\}\notag\\
&& \label{eq:L-dagger-exp-g}\\
\mathcal{L}^{\dagger}_{\textbf{s}}\left[e^{-i\alpha\textbf{k}\cdot\textbf{s}}\right] &=& e^{-i\alpha\textbf{k}\cdot\textbf{s}}\big\{-i\alpha(\nabla_{\textbf{s}}\cdot\textbf{D})\cdot\textbf{k}\notag\\
&& -\alpha^2\textbf{D}:\textbf{k}\textbf{k}^T+i\alpha\beta(\nabla_{\textbf{s}}U)^T\textbf{D}\textbf{k}\big\}\notag\\
&& \label{eq:L-dagger-exp}
\end{eqnarray}
It can be demonstrated that replacing the relevant terms in (\ref{eq:L-dagger-expand}) with (\ref{eq:L-dagger-g}), (\ref{eq:L-dagger-exp-g}) and (\ref{eq:L-dagger-exp}) and integrating with respect to $\textbf{r}$ leads to many cancellations. Eq.~(\ref{eq:appendix:Fxy-exp}) can therefore be simplified as:
\begin{eqnarray}
&&{\color{black}\int\int \mathcal{L}_\mathbf{s}^\dagger\left[F(\mathbf{s},\mathbf{r})\right] \delta(\mathbf{s}-\mathbf{r})p_0(\mathbf{r})\,d\mathbf{s}\,d\mathbf{r}} \notag\\
&& = \int d^3\textbf{s}\,p_0(\textbf{s})\bigg[\textbf{k}^T\textbf{D}(\textbf{s})\textbf{k}g(\textbf{s}-\textbf{r}_0)\notag\\
&&
-\frac{i}{\alpha}\textbf{k}^T\textbf{D}(\textbf{s})\nabla_{\textbf{s}}g|_{\textbf{s}-\textbf{r}_0}
\bigg]
\end{eqnarray}
Taking the real part yields the left-hand side of Eq.~(\ref{AverGammaC2}). 
{\color{black}
For the $\alpha\rightarrow0$ estimator of Eq.~(\ref{AverGammaC2-alpha0}), it can be noted that $F$ is given by:
\begin{eqnarray}
F(\mathbf{s},\mathbf{r}) &=& \tfrac14\left[\mathbf{k}\cdot(\mathbf{s}-\mathbf{r})\right]^2
\left[g(\textbf{s}-\textbf{r}_0)+g(\textbf{r}-\textbf{r}_0)\right]\notag\\
&& 
\end{eqnarray}
The gradient and Hessian of $F$ will thus be given by:
\begin{eqnarray}
\nabla_\mathbf{s}F(\mathbf{s},\mathbf{r}) &=& \tfrac12\mathbf{k}\mathbf{k}^T(\mathbf{s}-\mathbf{r})\left[g(\textbf{s}-\textbf{r}_0)+g(\textbf{r}-\textbf{r}_0)\right] +\notag\\
&& \tfrac14\left[\mathbf{k}\cdot(\mathbf{s}-\mathbf{r})\right]^2\nabla_\mathbf{g}(\mathbf{s}-\mathbf{r}_0)
\label{eq:grad-F-alpha0}\\
\mathbf{H}_{\mathbf{s},F}(\mathbf{s},\mathbf{r}) &=& \tfrac12\mathbf{k}\mathbf{k}^T\left[g(\textbf{s}-\textbf{r}_0)+g(\textbf{r}-\textbf{r}_0)\right]+\notag\\
&& \mathbf{k}\mathbf{k}^T(\mathbf{s}-\mathbf{r})\left[\nabla_\mathbf{s}g(\mathbf{s}-\mathbf{r}_0)\right]^T+\notag\\
&& \tfrac14\left[\mathbf{k}\cdot(\mathbf{s}-\mathbf{r})\right]^2\mathbf{H}_{\mathbf{s},g}(\mathbf{s}-\mathbf{r})
\label{eq:Hessian-F-alpha0}
\end{eqnarray}
Eqs.~(\ref{eq:grad-F-alpha0}) and (\ref{eq:Hessian-F-alpha0}) can then be used to obtain the full expression for $\mathcal{L}^\dagger_{\mathbf{s}}F(\mathbf{s},\mathbf{r})$. We, however, remark that such an expression will be integrated against $\delta(\mathbf{s}-\mathbf{r})$. As such, all the terms that contain $\mathbf{s}-\mathbf{r}$ will disappear. Therefore, the only surviving term will be the leading term in the Hessian, which will yield:
\begin{eqnarray}
	&&\lim\limits_{\tau\rightarrow 0^+}\frac{1}{\tau}\left\langle F(\mathbf{X}_{t+\tau},\mathbf{X}_t)\right\rangle_{\mathbf{X}_t\sim p_0(\mathbf{r})} = \notag\\
	&& \frac12\int\int \mathbf{D}: \tfrac12\mathbf{k}\mathbf{k}^T\left[g(\textbf{s}-\textbf{r}_0)+g(\textbf{r}-\textbf{r}_0)\right]\notag\\
	&& \times\delta(\mathbf{s}-\mathbf{r}) p_0(\mathbf{r})\,d\mathbf{s}\,d\mathbf{r} = \notag\\
	&& \int \mathbf{k}^T\mathbf{D}\mathbf{k} g(\mathbf{r}-\mathbf{r}_0)\,d\mathbf{r}
\end{eqnarray}}When it comes to small-- but nonzero-- $\tau$ values, one can employ a similar approach to obtain error terms using Eq.~\eqref{eq:appendix:Fxy-exp} {\color{black}for both estimators} as follows: 
{\color{black}
\begin{eqnarray}
	&&\int\int F(\mathbf{s},\mathbf{r}) \left\{\left[
	\frac{e^{\tau\mathcal{L}_\mathbf{s}}-1}{\tau}
	\right] \delta(\mathbf{s}-\mathbf{r})\right\}
	p_0(\mathbf{r})\,d\mathbf{s}\,d\mathbf{r}=\notag\\
	&&\int\int \left\{\left[
	\sum_{l=1}^{\infty}\frac{\tau^{l-1}\left(\mathcal{L}_{\mathbf{s}}^{\dagger}\right)^lF}{l!}
	\right]F(\mathbf{s},\mathbf{r})  \delta(\mathbf{s}-\mathbf{r})\right\}
	p_0(\mathbf{r})\,d\mathbf{s}\,d\mathbf{r} \notag\\
\end{eqnarray}
}
Given the above mathematical structure, it is possible to estimate the discretization error,~i.e.,~the terms arising due to the higher-order powers of $\mathcal{L}$. The most obvious (but not straightforward) means of doing so will involve applying $\mathcal{L}^\dagger$ to {\color{black}$F(\mathbf{s},\mathbf{r})$} multiple times and evaluating the integral. One might also consider solving the adjoint Smoluchowski equation numerically.\cite{HonischPhysRevE2011} It is also possible to derive higher-order finite difference approaches that allows for the usage of larger $\tau$'s and thus displacements over longer time windows.

\section*{ACKNOWLEDGMENTS}
\noindent A.H.-A. gratefully acknowledges the support from the National Science Foundation Grants CBET-1751971 (CAREER Award) and CBET-2024473. This work was supported as part of the Center for Enhanced Nanofluidic Transport (CENT), an Energy Frontier Research Center funded by the U.S. Department of Energy, Office of Science, Basic Energy Sciences under Award \#DE-SC0019112. These calculations were performed on the Yale Center for Research Computing. This work used the Extreme Science and Engineering Discovery Environment (XSEDE), which is supported by National Science Foundation Grant No. ACI-1548562.73.

\section*{References}

\bibliographystyle{achemso}

\bibliography{Diffusivity-Stochastic}

\end{document}